%% file: 00-sample-manuscript.tex
\documentclass[manuscript, nonacm]{acmart}

\AtBeginDocument{%
  \providecommand\BibTeX{{%
    \normalfont B\kern-0.5em{\scshape i\kern-0.25em b}\kern-0.8em\TeX}}}

\setcopyright{acmcopyright}
\copyrightyear{2025}
\acmYear{2025}
\acmDOI{XXXXXXX.XXXXXXX}

\acmJournal{JACM}
\acmVolume{37}
\acmNumber{4}
\acmArticle{111}
\acmMonth{8}

\acmPrice{15.00}
\acmISBN{978-1-4503-XXXX-X/18/06}

\usepackage{subcaption}
\usepackage{tikz}

\newcommand{\meatballicon}[2][1ex]{%
  \begin{tikzpicture}[baseline={(dot0.base)}, yshift=0.25ex] 
    \foreach \i in {0,1,2}{
      \coordinate (dot\i) at (\i*#1*2.5,0);
      \fill[#2] (dot\i) circle (#1);
    }
  \end{tikzpicture}%
}

\usepackage{multirow} 
\usepackage{soul}
\usepackage{multicol}
\usepackage{lipsum}
\usepackage{mwe}
\usepackage{array}
\usepackage{booktabs} 
\usepackage{multirow}
\usepackage{makecell}
\usepackage{threeparttable}
\usepackage{titlesec}
\titleformat{\subsubsection}[runin]{\normalfont\itshape}{\thesubsubsection.}{1em}{}[\newline]
\titlespacing*{\subsubsection}{0pt}{\baselineskip}{0pt}

\begin{document}


\title[A Critical Examination of the Accessibility of User-Enacted Moderation Tools on Facebook and X]{``I thought it was my mistake, but it’s really the design'': A Critical Examination of the Accessibility of User-Enacted Moderation Tools on Facebook and X}



\author{Sudhamshu Hosamane}
\email{sudhamshu.hosamane@rutgers.edu}
\affiliation{%
  \institution{Rutgers University}
  \city{New Brunswick, NJ}
  \country{USA}
  }

  \author{Alyvia Walters}
\email{alyvia.walters@villanova.edu}
\affiliation{%
  \institution{Villanova University}
  \city{Villanova, PA}
  \country{USA}
  }

\author{Yao Lyu}
\email{yaolyu@umich.edu}
\affiliation{%
  \institution{University of Michigan}
  \city{Ann Arbor, MI}
  \country{USA}
  }

\author{Shagun Jhaver}
\email{shagun.jhaver@rutgers.edu}
\affiliation{%
  \institution{Rutgers University}
  \city{New Brunswick, NJ}
  \country{USA}
  }

\renewcommand{\shortauthors}{Hosamane, Walters, Lyu and Jhaver}
\begin{abstract}
As social media platforms increasingly promote the use of user-enacted moderation tools (e.g., reporting, blocking, content filters) to address online harms, it becomes crucially important that such controls are usable for everyone.
We evaluate the accessibility of these moderation tools on two mainstream platforms -- Facebook and X -- through interviews and task-based walkthroughs with 15 individuals with vision impairments.
Adapting the lens of \emph{administrative burden of safety work}, we identify three interleaved costs that users with vision loss incur while interacting with moderation tools: \emph{learning costs} (understanding what controls do and where they live), \emph{compliance costs} (executing multi-step procedures under screen reader and low-vision conditions), and \emph{psychological costs} (experiencing uncertainty, stress, and diminished agency).
Our analysis bridges the fields of content moderation and accessibility in HCI research and contributes
(1) a cross-platform catalog of accessibility and usability breakdowns affecting safety tools; and (2) design recommendations for reducing this burden. 
\end{abstract}



\keywords{Content Moderation, Accessibility, Usability, Social Media Design}

\maketitle
\input{00-body}

\begin{acks}
 We are extremely grateful to our participants for their time, insights, and help recruiting others. We thank the New Jersey Commission for the Blind and Visually Impaired (CBVI) for sharing our initial recruitment call through their e-mailing list. We also thank Britt Paris for feedback and helpful discussions. This work was supported by the National Science Foundation under Grant No. 2329394. Any opinions, findings, and conclusions are those of the authors and do not necessarily reflect the views of the NSF or CBVI.
\end{acks}
\bibliographystyle{ACM-Reference-Format}
\bibliography{references}


\end{document}

%% file: 00-body.tex

\input{01-intro-v2}
\input{02-related-work-SH-v2}
\input{04-methods-v2}
\input{05-findings-v2}
\input{06-discussion-v2}
\input{07-conclusion}

%% file: 01-intro-v2.tex
\section{Introduction}






Online safety is a growing problem, encompassing challenges of persistent harassment and hate, fast-spreading misinformation, technology-facilitated privacy and sexual harms (e.g., non-consensual imagery), and deepfake-driven impersonation~\citep{Pew2021Harassment,Vosoughi2018FalseNews,HenryPowell2018TFSV,ChesneyCitron2019DeepFakes,FTC2024DataBook,scheuerman2018safe,wei2023there}. 
For marginalized populations especially, online safety is an increasingly concerning topic~\citep{kojah2025dialing,mayworm2024online,mayworm2024content}. For instance, prior research on the use of social media by \emph{people with visual impairments}\footnote{We adopt person-first terminology for visual disability as per our participants' preference.} has documented the online safety problems they encounter, including targeted attacks by trolls and harassment, and these problems are exacerbated by accessibility barriers and inadequate response from platform management \cite{lyu24blindtok, heung2024victim}. 

Social media platforms rely upon content moderation infrastructures to protect users from such harms and offer safety. Traditionally, platforms used third-party~\citep{Roberts2019} and volunteer moderators~\citep{Matias2019Civic,Wohn2019TwitchMods,McGillicuddy2016controlling} as well as automated tools~\citep{jhaver2019automod,chandrasekharan2019crossmod} to perform most moderation tasks, such as detecting and removing norm violations and sanctioning offenders~\citep{kiesler2012regulating}. 
However, platforms are increasingly delegating safety work to end users through \emph{user-enacted moderation} tools such as reporting, blocking, and content filtering~\citep{Crawford2016Flag, Jhaver2023Personalizing, jhaver2022filterbuddy, geiger2016bot}, and that delegation is intensifying. For example, recent announcements by Meta\footnote{Changes in policy enforcement to now rely more on user reports: https://about.fb.com/news/2025/01/meta-more-speech-fewer-mistakes/} signal greater dependence on user-submitted reports, leaving automated systems to prioritize only illegal or highest-severity content. Because of this shift in governance approach, it is crucial that \textit{all} users are able to report problematic content. In addition, given that the reliance on traditional approaches to moderation is eroding, it is vital that users are able to exercise adequate control over tools that remove the content they want to avoid.



Despite their utility, it is important to consider that the application of user-enacted moderation tools is complex. Prior research has documented the time, cognitive demands, and careful social considerations that controlling and configuring these tools requires on the part of the end-user ~\citep{Crawford2016Flag,Jhaver2023Personalizing,jhaver2018blocklists}. For users with visual impairment, the operational accessibility of user-enacted content moderation tools is of still greater concern.
This is because a substantial body of HCI and accessibility research shows that even \emph{basic}, single-step interactions on mainstream social media—discovering menus, interpreting icon-only buttons, and reading posts—can be inconsistent or difficult to perform non-visually~\citep{Jordan2024wayfinding,Ross2018buttons,morris2016twitterbarelyuse}. Against this backdrop, we argue that inaccessible moderation controls undermine user safety, impose disproportionate interactional burdens~\citep{Branham2015InvisibleWork}, and deny people with vision-related disabilities from attaining the agency that the use of these tools promises.


Recent estimates indicate that nearly \emph{43 million} people are blind worldwide, with \emph{hundreds of millions} more living with low vision~\citep{WHO2019WorldReport,GBDVision2021}. Moreover, prior research establishes that social media platforms are vital social lifelines for this population. For example, studies of Facebook use have found that blind users are often highly active participants who build larger and more diverse social networks than their sighted peers, using the platform as a key resource for social support~\citep{wu2014blvFB}. Thus, ensuring their safety is vital to preserving their access to these essential digital spaces. Further, solutions created for users facing the greatest barriers often reveal and fix fundamental design flaws, leading to more robust and intuitive systems for all users~\citep{Rigot2022margins}.

In this article, we examine the accessibility of user-enacted moderation tools through in-depth, task-based interviews with 15 social media users with visual impairments on Facebook and X (formerly Twitter). 
Rather than just asking whether these controls are \emph{perceivable} or \emph{usable}, we center \emph{operational accessibility}: whether users with blindness or low-vision (BLV) can \emph{complete} safety actions end-to-end.
Our analysis shows how current designs obscure or complicate essential controls across the entire workflow: locating an entry point, navigating multi-step pages with inconsistent layouts, understanding the consequences of actions, and verifying outcomes that are often hidden. We frame these breakdowns using the lens of \emph{Administrative Burden}—a frame which allows us to distinguish \emph{learning}, \emph{compliance}, and \emph{psychological} costs—to articulate how small frictions at any stage compound and propagate, making later steps harder and eroding user confidence.  

This paper extends HCI literature at the intersection of accessibility, online safety, and platform governance. We provide a crucial empirical contribution: a rich, cross-platform catalog of the specific interactional breakdowns and navigational failures that users with vision loss experience when attempting to moderate content on Facebook and X. Grounded in these lived experiences, we also provide design recommendations to make present-day moderation controls more equitable and controllable. We accomplish this by importing and adapting the concept of Administrative Burden as a novel analytical framework to examine the issues in operational accessibility of user-initiated safety workflows. This lens moves beyond standard usability evaluations to provide a precise vocabulary—learning, compliance, and psychological costs—for articulating uncertainty and labor that inaccessible safety tools offload onto marginalized users.


%% file: 02-related-work-SH-v2.tex
\section{Related Work}
\label{sec:related_work}


We first provide an overview of \emph{content-moderation research on social-media platforms}, highlighting its focus on user-enacted moderation tools and an assumed ``sighted default'' in their design (Section~\ref{ref:related_work:1}). We then survey \emph{digital accessibility work}, cataloging the barriers that users who are blind or have low-vision (BLV) face in routine social media tasks (Section~\ref{ref:related_work:2}). Synthesizing these strands reveals an under-examined problem—the inaccessibility of safety tools for users with visual disabilities—that our study directly tackles.

\subsection{Overview of User-Enacted Content Moderation on Social Media}
\label{ref:related_work:1}

Social media platforms bear the key responsibility of shaping online safety by defining and enforcing community standards of appropriate conduct~\citep{riedl2021responsible,helberger2018governing}. 
In enforcing these standards, they identify and regulate a wide array of problematic content, from illicit material and coordinated harmful behavior to hate speech, misinformation, and interpersonal harassment~\citep{scheuerman2021framework}.
Platform regulation is critical given that such norm violations can profoundly deteriorate user well-being, public discourse, and the very integrity of these digital environments~\citep{Napoli2019social,Gillespie2018}. Most efforts to address these issues have relied on centralized, platform-driven interventions. These typically involve a dynamic interplay between increasingly sophisticated automated detection systems designed to flag inappropriate content at scale and vast teams of professional human moderators tasked with adjudicating policy violations, especially those that are too nuanced or unclear to be judged by automated systems~\citep{Gillespie2018, Roberts2019}. While this mixed initiative approach is necessary to handle the large volume of content, it presents problems such as the potential for algorithmic bias and the significant, well-documented psychological burden placed upon human reviewers who are incessantly exposed to distressing material~\citep{Roberts2019, Steiger2021moderator, Matias2019Civic}.

Recognizing the limitations of purely centralized models, platforms are increasingly delegating the responsibility of content governance to end-users. This includes community-led moderation (e.g., by end-users who serve as volunteer moderators on Reddit or in Facebook Groups)~\citep{Matias2019Civic, Seering2020selfmoderation} and, more pervasively, the provision of individualized user-driven moderation tools~\citep{Jhaver2023Personalizing}. Features like blocking, muting, and reporting content are now standard across platforms, aiming to empower individuals to tailor their experiences and flag harmful content for platform review~\citep{jhaver2018blocklists, Crawford2016Flag, Gillespie2018}.
 Scholarly inquiry into user-enacted moderation has explored a variety of dimensions from the perspective of the general user base, including the motivations behind using these tools (i.e., why users block~\citep{jhaver2018blocklists} or report~\citep{zhang2023cleaning} others), the perceived fairness and effectiveness of reporting systems~\citep{shim2024incorporating}, and user preferences regarding the use of personalized content filters versus broader platform-level interventions to address norm violations~\citep{Jhaver2023Personalizing}.

Most closely related to our work, Heung et al.\ have explored how configurable AI-assisted filters could protect disabled users from ableist hate, highlighting the need for granular, user-controlled safety mechanisms~\citep{heung2025ignorance}. Our study extends this agenda by asking whether users with visual impairment can even \emph{access} the baseline moderation tools—an essential prerequisite before such tools can be useful.

On the whole, a critical examination of extant research on user-enacted moderation reveals a pervasive \emph{``sighted default''}: an implicit assumption that moderation tools are designed for and used by people who can see. For example, literature on the use of reporting, blocking, and other safety tools overwhelmingly assumes a legibility of visual elements like icons, menus, and layouts, while largely ignoring whether these crucial instruments are accessible to users with disabilities. In particular, the challenges faced by individuals with vision impairments who use screen readers remain significantly under-addressed. We argue in this article that this oversight is not trivial---the functional accessibility of moderation tools is a prerequisite for equitable participation in platform governance. It represents a foundational research gap that leaves a substantial user base unprotected and excluded from a core aspect of digital citizenship.

\subsection{Accessibility in Social Media: Persistent Barriers for Users with Visual Disabilities}
\label{ref:related_work:2}

Social media platforms like Facebook and X are integral to contemporary life for many, including those with vision-related disabilities~\citep{lowTwitterA11yBrowser2019,gleasonTwitterA11yBrowser2020,gleasonAddressingAccessibilitySocial2019}. These users were early adopters, benefiting from the platforms' initial text-based nature~\citep{Voykinska2016SNS, wu2014blvFB}, and continue to engage heavily with them~\citep{Gkatzola2024review}. Yet widespread accessibility barriers persist—not by accident, but due to systemic neglect of core user-centered design principles~\citep{Wobbrock2011Ability, norman2013doet}. In particular, violations of the Web Content Accessibility Guidelines (WCAG)—Perceivability, Operability, Understandability, and Robustness (POUR)~\citep{wcag22}—create frequent usability breakdowns for users with visual impairments.

The most fundamental barriers undermine the \textit{``Perceivable''} and \textit{``Operable''} principles of WCAG. Visual content such as images and memes often lacks appropriate alternative text, rendering them incomprehensible to screen readers~\citep{Gleason2020Memes}. Automated captioning systems exist, but their accuracy and context awareness remain limited, so they have not fully solved this problem~\citep{stangl2020aatnocontext}. Critically, interactive elements like buttons and menus often lack proper semantic labels. This forces screen reader users into laborious trial-and-error navigation for basic tasks~\citep{morris2016twitterbarelyuse}, as they often cannot perceive the function of a control or operate the interface effectively.

Users with low-vision face a different, yet equally challenging, set of barriers. While they may use aids like virtual screen magnifiers, interfaces are often not \textit{``Robust''} enough to handle them gracefully. Magnification can disrupt page layouts, requiring tedious panning that obscures page context and violates basic usability tenets~\citep{szpiro2016blviAccess, tang2023}. Further, when accessibility-based gestures conflict with app-based gestures or become confusing to track, it makes routine actions cognitively taxing and error-prone~\citep{bennett2018instagram}. 

These persistent accessibility gaps in everyday social media use suggest a grim prognosis for the usability of a broader range of tools. When basic tasks like reading a feed, following links, and manipulating buttons remain difficult for users with vision loss \citep{lyuBecauseSightedPeople2024,lyuUploadAllTypesDifferent2024}, it is reasonable to expect that more complex, multi-step features such as reporting flows or other trust-and-safety tools will be even less accommodating. Independent accessibility audits support this concern: investigators have documented unlabeled controls and other flaws that make moderation workflows effectively unusable with screen readers~\citep{thiel2022}. While such audits begin to offer valuable initial insights into these challenges, they were brief spot-checks. As such, they stopped at listing missing labels—without tracking participants across multi-screen reporting process, measuring task-completion drop-offs, or examining other subtler obstacles that users with vision impairment could face. Broader analyses of accessibility research trends suggest that nuanced investigations into complex, safety-critical workflows like content moderation for users with vision-related disabilities are still relatively scarce compared to studies on more foundational interaction or consumption tasks~\citep{JeanneretMedina2025TACCESS}. 
 


\vspace{12pt}
 \noindent This recurring pattern of inaccessibility in safety-critical tools extends beyond isolated usability flaws; it reflects a systemic failure to apply principles of design justice and critical disability studies~\citep{CostanzaChock2020, Spiel2020Nothing,Sum2022Dreaming}. This oversight effectively violates the disability rights imperative of ``\emph{Nothing About Us Without Us}''~\citep{charlton1998nothing}. When safety tools are not designed inclusively, users with vision-related disabilities are forced to perform significant ``\emph{invisible adaptation work}'' just to achieve basic safety~\citep{Branham2015InvisibleWork}, a burden likely intensified for those who may already face disproportionate online harm~\citep{heung2022microaggressions, heung2024victim}.

 Emerging studies are beginning to highlight the tangible consequences of this accessibility gap. For instance, Lyu et al. found that TikTok users who were blind struggled to report harassment due to inaccessible interface elements, and some ended up getting unfairly sanctioned as a result~\citep{lyu24blindtok}. Further, a recent examination of misinformation warnings revealed that many participants experiencing visual impairment did not notice platform-issued warning labels on YouTube, Facebook, and Tiktok, indicating those visual cues were effectively invisible to them~\citep{sharevski2023blindmisinfo}. 

Our research builds upon these early findings by conducting a systematic, in-depth examination of how users with vision impairments experience and interact with prevalent user-enacted moderation tools. 
By centering these users' moderation needs, we seek to examine the present-day accessibility of user-initiated moderation tools. 
This investigation is guided by the following research questions:

\begin{enumerate}
\item \textbf{RQ1:} How do blind and low-vision users currently perceive, navigate, and attempt to utilize common user-enacted moderation tools (e.g., reporting, blocking, content filtering) available on mainstream social media platforms?

\item \textbf{RQ2:} What specific usability and accessibility barriers do users with vision impairment encounter when engaging with these moderation interfaces, and how do these barriers impact their ability to manage their safety and online experience effectively?

\item \textbf{RQ3:} What design improvements and systemic changes are needed to create genuinely accessible and equitable user-enacted moderation tools for people with visual impairments?
\end{enumerate}

%% file: 04-methods-v2.tex
\section{Methods}
\label{sec:methods}

\subsection{Study Participants and Protocol}
\label{sec:methods:1}

This study employed a qualitative methodology to investigate the experiences of individuals with blindness or low-vision (BLV) with user-enacted content moderation tools. We studied tools on two major social media platforms: Facebook and X. Our approach centered on observing participants as they performed a series of representative moderation, privacy, and safety-related tasks on their own mobile devices. 
Ethical approval for all study procedures was obtained from the Institutional Review Board (IRB) at [University name will be disclosed after the peer review completes].

We recruited fifteen participants from New Jersey who were 18+ years old, identified as having BLV, were regular Facebook or X users, proficient with assistive technology, and were fluent in English. Recruitment proved challenging though ultimately successful; after unproductive online advertising and unsuccessful outreach to local organizations and universities, participants were sourced through a community mailing list and subsequent word-of-mouth referrals.

Data collection occurred between November 2024 and April 2025, with the majority of interviews (13 out of 15) conducted between February and April 2025. To enhance ecological validity and participant comfort, all study sessions were conducted in person and at locations chosen by the participants, which included their homes, local libraries, or quiet coffee shops. Each participant used their own smartphones\footnote{All participants owned an iPhone and used it as their primary device for performing moderation tasks during the interview. They were given the option to use any digital device they were comfortable with.}, their preferred screen reader software (all 15 participants in this cohort used VoiceOver on iOS), and their preferred social media platform (six used X, the rest opted for Facebook). This ensured that interactions took place within a familiar technological setup. Sessions were comprehensive, lasting between 2 hours and 5 minutes and 3 hours and 15 minutes (median duration: approximately 2 hours and 15 minutes), allowing sufficient time for rapport-building, task completion, and a debriefing at the end. Participants were encouraged to take adequate breaks in between tasks.  

\begin{table}[ht]
\centering
\caption{Participant Demographics and Media Usage Background}
\label{tab:participants}
\begin{tabular}{l c c l l l}
\toprule
\textbf{Pseudonym} & \textbf{Age} & \textbf{Platform} & \textbf{Condition} & \textbf{VoiceOver (yrs)} & \textbf{Platform (yrs)} \\
\midrule
Shreya & 21 & X & High myopia + retinal issues & 5 & 13 \\
Kiara & 25 & X & Leber Congenital Amaurosis & 12 & 12 \\
Samuel & 26 & Facebook & Retinal detachment (lights only) & 10+ & 14 \\
Jamila & 30 & Facebook & Congenital glaucoma (light only) & 14 & 15 \\
Maya & 35 & X & Pseudotumor Cerebri (light only) & 6 & 13 \\
Adriana & 36 & Facebook & Retinitis Pigmentosa (RP) (legally blind) & 15 & 16 \\
Jessica & 36 & Facebook & Congenital glaucoma (total blind) & 11 & 13 \\
Grant & 38 & X & Glaucoma + Aniridia (light only) & 9 & 5 \\
Mateo & 45 & X & Glaucoma (light perception in one eye) & 4.5 & 14 \\
Cristian & 51 & X & Glaucoma (totally blind since 17) & 13 & 6 \\
Frank & 57 & Facebook & RP + Stargardt (light only) & 6 & 15 \\
Nicole & 62 & Facebook & RP (legally blind, tunnel vision) & \textless1 & 15 \\
Laverne & 63 & Facebook & Glaucoma + Failed Corneal Transplant & 12 & 1 \\
Vasu & 66 & Facebook & Glaucoma (total blind) & 15 & 13 \\
Patricia & 81 & Facebook & Optic Neuritis, RP (total blind) & 13 & 15 \\
\bottomrule
\end{tabular}
\end{table}

A detailed summary of participant demographics, including pseudonyms, age, social media platform used during the study, self-reported visual condition, and years of experience with VoiceOver and the respective platform, is provided in Table~\ref{tab:participants}. The cohort ranged in age from 21 to 81, with most having extensive experience (median of approximately a decade) using screen readers and mobile technology. Thirteen participants reported having complete blindness or having only minimal light perception, while the remaining two had severe low vision. It is crucial to underscore, however, that the lived realities of these individuals are far more nuanced than a table can fully capture. Participants experienced vision loss at different stages of life—some were blind since birth, others experienced significant vision loss later in life (e.g., after age 60), and some were in the ongoing process of losing their sight. Consequently, their journeys with technology varied: some learned to use social media for the first time after vision loss, while others had to relearn interaction paradigms using a screen reader after years of sighted use. 

\subsection{Task Design and Administration}

To investigate how users with visual disabilities engage with content moderation controls and any specific barriers they might encounter, we designed a sequence of \emph{six} tasks. These tasks were based on the variety of content moderation controls (previously discussed in Section \ref{ref:related_work:1}) available on both Facebook and X—the two social media platforms used in this study—and were meant to be evaluated against established interaction-design principles such as discoverability, feedback, consistency, and user control~\citep{norman2013doet, shneiderman2017dui, Nielsen1994}, which align with core tenets of the Web Content Accessibility Guidelines (WCAG)~\citep{WAI2023, wcag22}. The tasks progressed from basic content interaction and reporting to more specific safety and privacy actions, reflecting common user-oriented moderation activities.

\subsubsection{Task Environment and Stimuli}

To ensure a standardized yet realistic context for task performance while prioritizing participant privacy and avoiding unintended consequences for their personal social media accounts, all tasks were conducted using controlled account environments.

For interactions on X, two study-specific dummy accounts were created by the research team: `Jenny Chang' and `Marcus Chang'.\footnote{These pseudonyms were randomly generated and do not represent real individuals; realistic-sounding names were chosen to ensure we weren't blocked from the platform.} The `Marcus Chang' account was used to post simulated problematic content directed at `Jenny Chang's' profile by tagging the latter's username. To avoid exposing participants to distressing stimuli, this content often included placeholders (e.g., ``[placeholder for hateful slur]'' or ``[placeholder for misinformation claim about an event]'') rather than displaying actually offensive material. Participants were informed of the purpose of these messages and placeholders. During these sessions, we helped participants login into the dummy account of `Jenny Chang', and instructed them: ``Imagine you are Jenny Chang, and you've encountered these posts from Marcus Chang on your profile. If you were in her shoes, how would you proceed doing \textit{x} (the tasks to follow)?'' We followed an equivalent procedure on Facebook.\footnote{Facebook’s strict scrutiny of fake accounts prevented creation of dummy profiles, so we used the first author’s rarely used personal account (with a limited social graph) as the recipient and a collaborator’s low-activity account as the sender. The collaborator posted placeholder content that tagged the author; visibility of the posted messages was restricted to just the two accounts, mirroring the X setup while safeguarding participant privacy.} 
%
At the end of the interview, we helped participants log out of the study accounts and log back into their original accounts.

\subsubsection{Moderation Tasks and Rationale}

We asked participants to complete the following tasks:

\begin{enumerate}

  \item \textbf{Report a Hateful Post:} Participants encountered a simulated post containing hostile or discriminatory content and identified the platform's flagging tool in order to report the post. We designed this task to assess the initial \emph{discoverability}~\citep{norman2013doet} and \emph{navigability} of flagging procedures through different context menus until they completed the task.


  \item \textbf{Report a Dummy Spam Group/Community:} We asked participants to navigate to a pre-created, inactive fake group (Facebook) or community (X) and report it for spam. This task broadened the scope from individual posts to a group, enabling us to evaluate different reporting contexts for \textit{consistency}~\citep{shneiderman2017dui}.

  \item \textbf{Block a User:} We instructed participants to block the account that was posting problematic content. This simulation provided a direct response to harassment and allowed us to evaluate the process of enacting interpersonal restrictions, a key aspect of \emph{user control and freedom}~\citep{Nielsen1994}.
  
  \item \textbf{Change Post Visibility Settings:} We asked participants to modify the audience of a sample post (i.e., from public to friends-only). This task explored participants' ability to manage content privacy, reflecting \emph{user control} over their own information dissemination.

  \item \textbf{Unblock the Same User:} After completing the account-blocking task, we asked participants at a later stage to locate and reverse the blocking action. This tested the \emph{reversibility} of actions~\citep{norman2013doet} and \emph{ease of redressal}, crucial for managing dynamic social boundaries.
  
  \item \textbf{Adjust Sensitivity Settings:} Both Facebook and X offer content filtering (e.g., for political or sensitive content); we asked participants to locate and modify these preferences to see `less' of this content (from the previously set `default' level). This task allowed us to examine the accessibility of controls related to \emph{personalized content experiences} and \emph{proactive filtering}.   
  \end{enumerate}

  \subsubsection{Think-Aloud Protocol and Data Capture}

   Throughout the task-based portion of the session, we employed a combination of \emph{concurrent} and \emph{retrospective} think-aloud protocols~\citep{Savva2015ConcurrentRetrospective}. Participants were encouraged to verbalize their thoughts, actions, and expectations as they interacted with the interfaces. We accounted for expert screen-reader users' tendency to listen at very high speech rates and to process information rapidly. This intense focus can sometimes lead to participants forgetting to verbalize their thought processes concurrently. Therefore, to ensure rich data capture, interviewers provided gentle, periodic nudges such as ``What is VoiceOver saying now?'' or ``What are you trying to do on this screen?'' to encourage continuous narration, especially during extended silences.

  Following each task, or at points of significant struggle, we asked brief retrospective questions (e.g., ``What were you searching for there?'' or ``Can you try to remember what you found difficult about that step?'') to clarify participants' mental models and strategies. We phrased questions in simple language and provided ample time for task completion and screen reader use. The interviewer minimized direct observation of the screen when it might disrupt the participant's auditory focus, relying primarily on the verbalized think-aloud data. All sessions were audio-recorded and later transcribed verbatim for analysis.

  \subsection{Data Analysis}
We employed reflexive thematic analysis in line with Braun and Clarke's guidelines \citep{Braun2006}. The first author transcribed all the audio recordings to text using Assembly AI,\footnote{https://www.assemblyai.com} verified a part of each transcript for correctness, and  conducted an exhaustive, line-by-line open-coding~\citep{strauss2015basics} of every transcript. This open-coding generated 155 preliminary codes that captured salient behaviors and reflections linked to our research questions. To strengthen rigor, the second author independently open-coded a strategically selected subset of transcripts. We then compared our respective code lists, negotiated differences, merged overlapping labels, and produced a consolidated codebook for subsequent iterative coding using this shared codebook as a scaffold. Next, the full author team engaged in axial coding—iteratively clustering related codes, refining category boundaries, and articulating higher-level patterns. Through successive rounds of discussion and memo-writing, we converged on five overarching themes that best synthesized participants' experiences and insights: \emph{Discoverability} (how easily participants could find the entry access point to proceed navigating to the moderation features), \emph{Navigability} (ease of moving within the interface elements and finally completing the task), \emph{Adaptation and Agency} (users' ability to control settings and customize experiences), \emph{Cognitive Load} (mental effort required, including challenges of memory and attention), and \emph{Expectation Mismatches} (discrepancies between what users expected and what the interface provided). We used spreadsheets and qualitative data software to organize codes and evidence (i.e., Excel for coding matrices and NVivo for annotating transcripts). 

\subsection{Researcher Positionality}
All members of the research team are sighted, and the first author led the interview and analysis work with the guidance of the other three. We acknowledge that this positionality influences our approach: for instance, we may not immediately notice accessibility issues that users with BLV experience daily. To address this, our team included advisors with expertise in BLV accessibility and content-moderation systems. We also consulted informally with other colleagues who work on accessibility and disability justice to review our procedures and interpretations. 
The second and fourth authors are specialists in user-centered social media governance research, while the third author is an expert in accessible technology evaluation; their prior experiences shaped our study design. 
We approached our analysis reflexively, maintaining an active commitment to disability-centered design principles by prioritizing participants' own descriptions and framing of their experiences and actively questioning our interpretations to ensure they faithfully represented the participants' perspectives. 

%% file: 05-findings-v2.tex
\section{Findings}
\label{sec:findings}

Our findings reveal a pervasive pattern of accumulated design shortcomings within Facebook and X's moderation interfaces that impose a significant burden on users with vision loss. We frame these results as the \emph{administrative burden of safety work}—the \emph{learning}, \emph{compliance}, and \emph{psychological} costs people incur when they try to protect themselves online. Learning costs are the effort required to figure out what controls exist, where they are, and how they behave; compliance costs are the effort required to carry out multi-step procedures correctly (gestures, detours, restarts); and psychological costs are the frustration, stress, anxiety, and loss of agency that arise when outcomes are opaque or tools do not work as intended. We borrow this triad from public administration scholarship on administrative burden~\citep{moynihan2015administrative} and apply it to end-user safety tasks in consumer platforms, where burdens appear as frictions in service encounters\footnote{Originating in citizen–state contexts, administrative burden denotes the costs individuals incur when engaging with policies and procedures, such that their implementation feels onerous. In that literature, \emph{learning costs} are the work of becoming aware of a program, understanding eligibility and benefits, and figuring out how to apply and maintain access; \emph{compliance costs} are the time and effort to complete required steps (forms, documentation, waiting, travel); and \emph{psychological costs} are the stress, frustration, stigma, and loss of autonomy that accompany these encounters. We transpose this typology to platform-administered safety workflows, treating user protection controls as service encounters, and use it as an analytic lens to locate where non-visual design choices create or relieve burden.}. As we detail in this section, our data suggest that these costs arise at different stages of the safety-task flow and, at times, cascade, with early frictions followed by greater effort and uncertainty in subsequent steps.

Although all 15 participants were active Facebook/X users comfortable with VoiceOver, only two had ever used reporting or blocking; most knew these features existed but not where to find them, reflecting low awareness and little prior need. This widespread inexperience meant that participants approached the moderation tasks as novices.


We begin with discovery and navigation hurdles (Section \ref{sec:findings:disc_nav}), where learning costs (finding/understanding entry points) and compliance costs (multi-step flows, detours, restarts) accumulate. We then show how design inconsistencies (Section \ref{sec:findings:inconsistencies}) add learning costs (misaligned mental models) and psychological costs (doubt, second-guessing). Finally, we trace consequences (Section \ref{sec:findings:labor_agency}): workarounds and rework that raise compliance costs, alongside emotional labor and diminished agency as psychological costs.

\subsection{Discovering and Navigating Safety Controls}
\label{sec:findings:disc_nav}

Participants usually navigate digital devices with a screen reader like VoiceOver. On each screen, Voiceover typically focuses on the first UI (user interface) element.\footnote{VoiceOver visually indicates its current focus with a black rectangle around the focused item (as shown in Figures~\ref{fig:back_focus} and~\ref{fig:more_focus}).} 
Interaction relies on a sequence of gestures: linear navigation (swiping left/right to move between elements sequentially), touch exploration (dragging a finger to identify elements by location), and custom actions (e.g., swiping up/down on an element to reveal context menus—akin to a long press). Effective navigation thus depends entirely on audible cues from properly labeled elements (buttons, links, text, etc.) and a predictable interface sequence.

Successfully using any moderation tool—the task we continually asked participants to undertake during our sessions—unfolds in two critical stages for all users, including users with BLV. First, users must \emph{discover} the initial entry point—typically an icon or menu link that houses controls like ``Report'' or ``Block.'' Second, they must \emph{navigate} the subsequent multi-step workflow of panels and selection screens to complete the action. A failure at either stage can render the entire feature unusable. Accordingly, we structure our findings to follow this user journey, first examining the barriers to discoverability and then the challenges of workflow navigability. 

\subsubsection{Discoverability Challenges} 
A recurring theme across our interviews was the difficulty participants faced in \emph{locating the initial entry point}—typically an icon or button—that would allow them to begin navigating the sequence of options required to complete a moderation task. One particularly problematic element was the small, often low contrast ``\raisebox{0.5ex}{\meatballicon[0.4ex]{lightgray}}'' menu icon—commonly known as the meatball menu\footnote{We refer to the ubiquitous meatball menu icon, used on both Facebook and X, to group additional actions, simply as the menu icon henceforth.}. This icon often houses crucial moderation options such as reporting a post, blocking a user, or flagging a group. Discoverability failures stemmed from two main issues: \emph{counterintuitive focus order} for screen reader users and \emph{poor visual contrast} of menu icons for users with low-vision.

\begin{figure}[ht]
  \centering
  \begin{subfigure}[b]{0.48\linewidth}
    \centering
    \includegraphics[width=\linewidth]{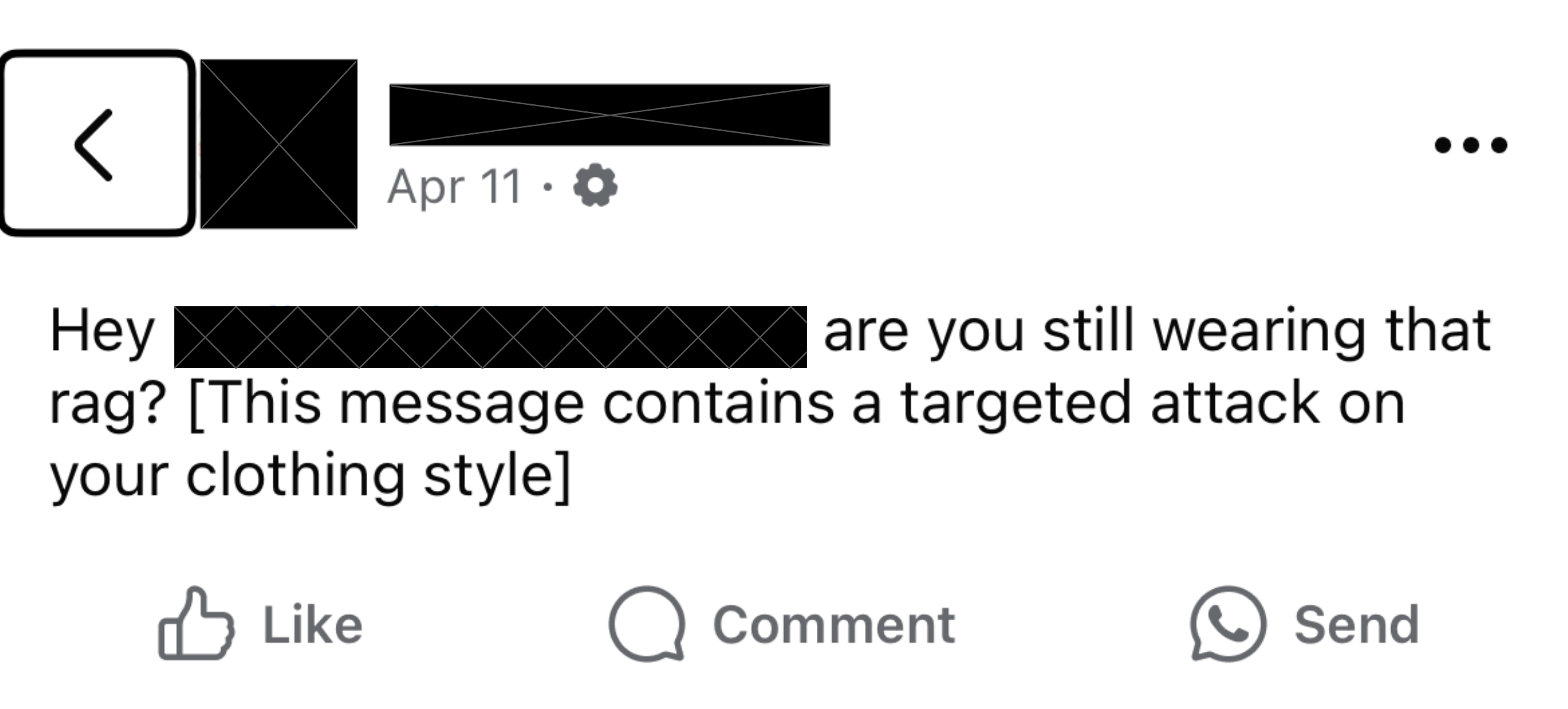}
    \caption{VoiceOver focuses on the ``Back'' button (top‐left) by default.}
    \label{fig:back_focus}
  \end{subfigure}
  \hfill
  \begin{subfigure}[b]{0.48\linewidth}
    \centering
    \includegraphics[width=\linewidth]{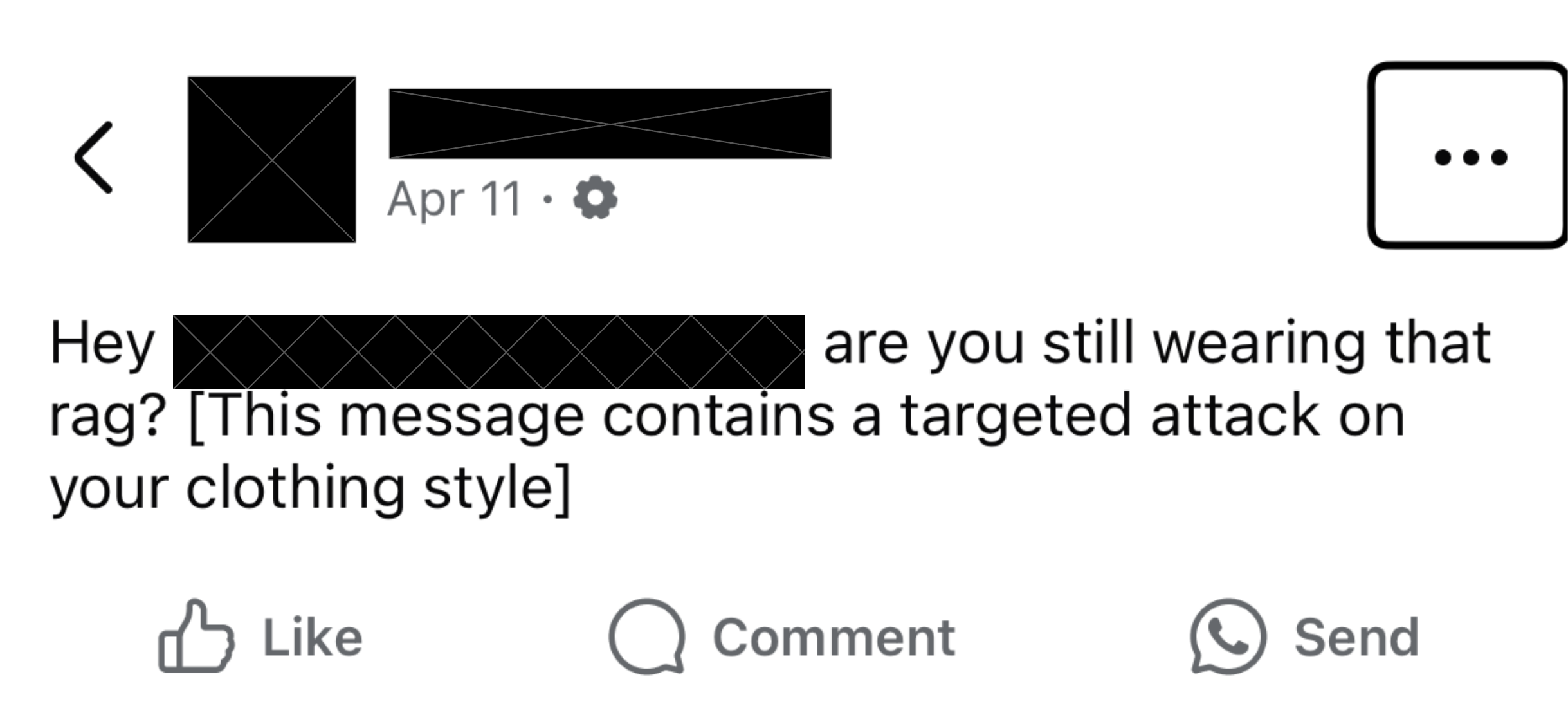}
    \caption{VoiceOver moves focus to the ``More options'' menu icon (top-right) after swiping left (\emph{not right}) from ``Back.''}
    \label{fig:more_focus}
  \end{subfigure}
  \caption{Misleading focus order in Facebook's interface}
  \label{fig:focus_comparison}
  \Description{Two side-by-side iOS screenshots of a Facebook post. Left image: focus is on Back (button) at the top left; the post shows a redacted profile, date, and bottom actions (Like, Comment, Send). A three-dot menu is visible at the top right. Right image: focus is on the More options (three-dot) button at the top right; the rest of the post layout (redacted profile, date, Like/Comment/Send) is the same as the left image.}
\end{figure}

When Jessica attempted to report an offensive Facebook post, she opened it and began methodically exploring its layout for the reporting control. VoiceOver announced other elements within the post, such as the author, time-stamp, and Like/Comment controls, but the ``More Options'' menu icon, which is the entry point for reporting, was never announced in this expected sequence (Figure~\ref{fig:focus_comparison}). \emph{``It didn't speak it to me... I had to go out of the post to get it,''} she explained. Frustrated, she had to navigate back to the main news feed—the list of all posts—where she could access a different context menu on the post summary that finally revealed the report function.

Prompted by the sighted interviewer, who confirmed the menu icon was visually present at the top-right of the expanded post view, Jessica returned to investigate and found that the focus order was reversed. She found that the ``More Options'' menu was only accessible with VoiceOver by swiping \emph{left} from the `Back' button—a counter-intuitive sequence, as screen reader users expect right swipes to move focus forward through a screen's layout, typically from left to right. Vasu encountered the same anomaly, calling it \emph{``backwards from how the screen looks.''}
%
%

Low-vision participants often knew where moderation controls should be but struggled to locate them. On X, reporting tools are placed in a pale white menu icon that blends into the interface and does not scale with iOS text-size settings. When Shreya—who has severe myopia and uses iOS Display Zoom (to increase text size)—was asked to report a tweet, she tapped the upper-right corner ``on muscle memory,'' yet admitted: \emph{``I can’t see the dots unless I hold the phone right up to my face... they're not really visible, it's like a little faded gray, not completely in your face.''} Even with Zoom-in enabled, the icon stayed small and washed-out, leading her to remark: \emph{``I think even someone with perfect sight might miss it at first glance.''}

\begin{figure}[ht]
  \centering
  \begin{subfigure}[b]{0.48\linewidth}
    \centering
    \includegraphics[width=\linewidth]{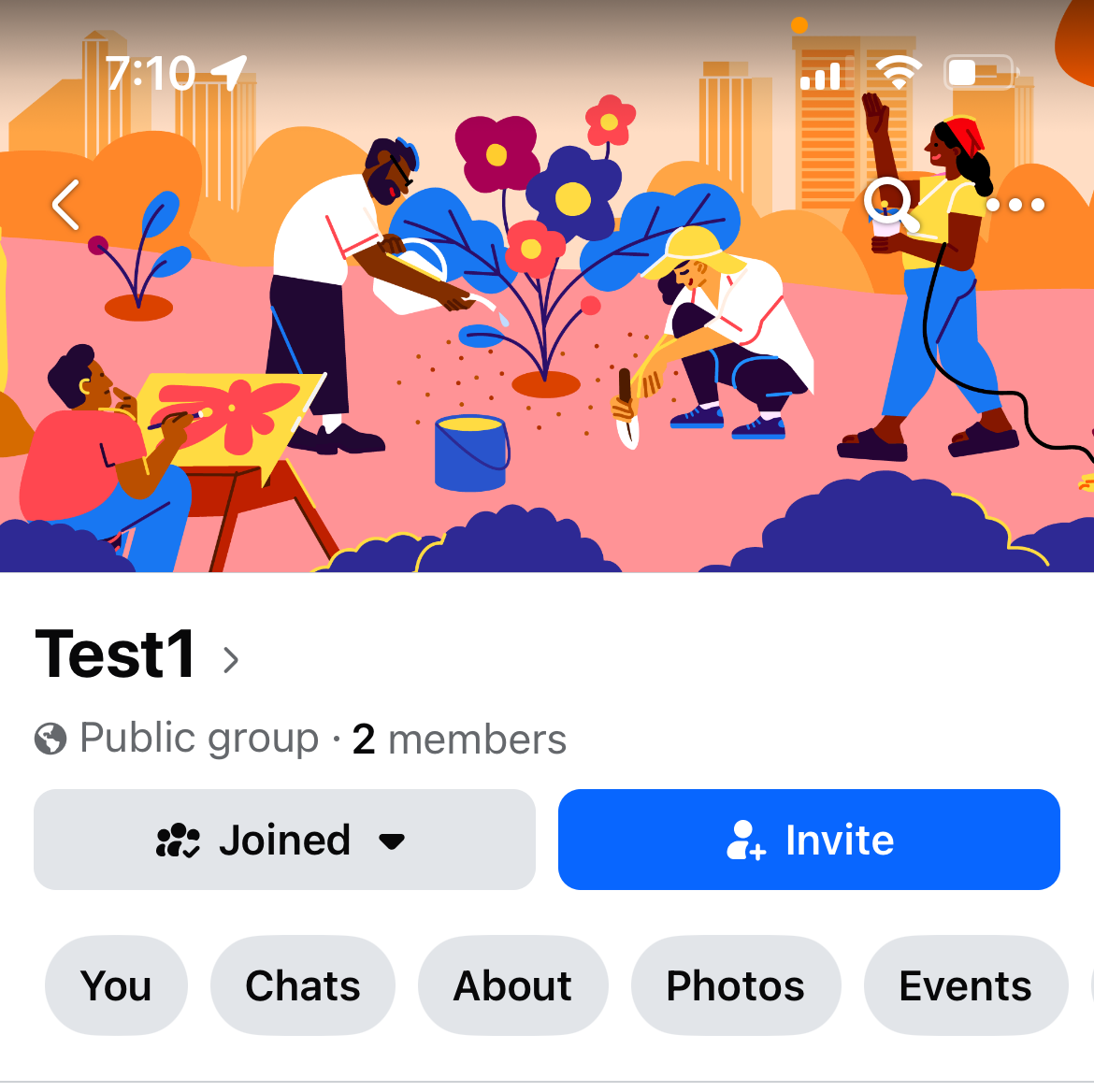}
    \label{fig:cover1}
  \end{subfigure}
  \hfill
  \begin{subfigure}[b]{0.48\linewidth}
    \centering
    \includegraphics[width=\linewidth]{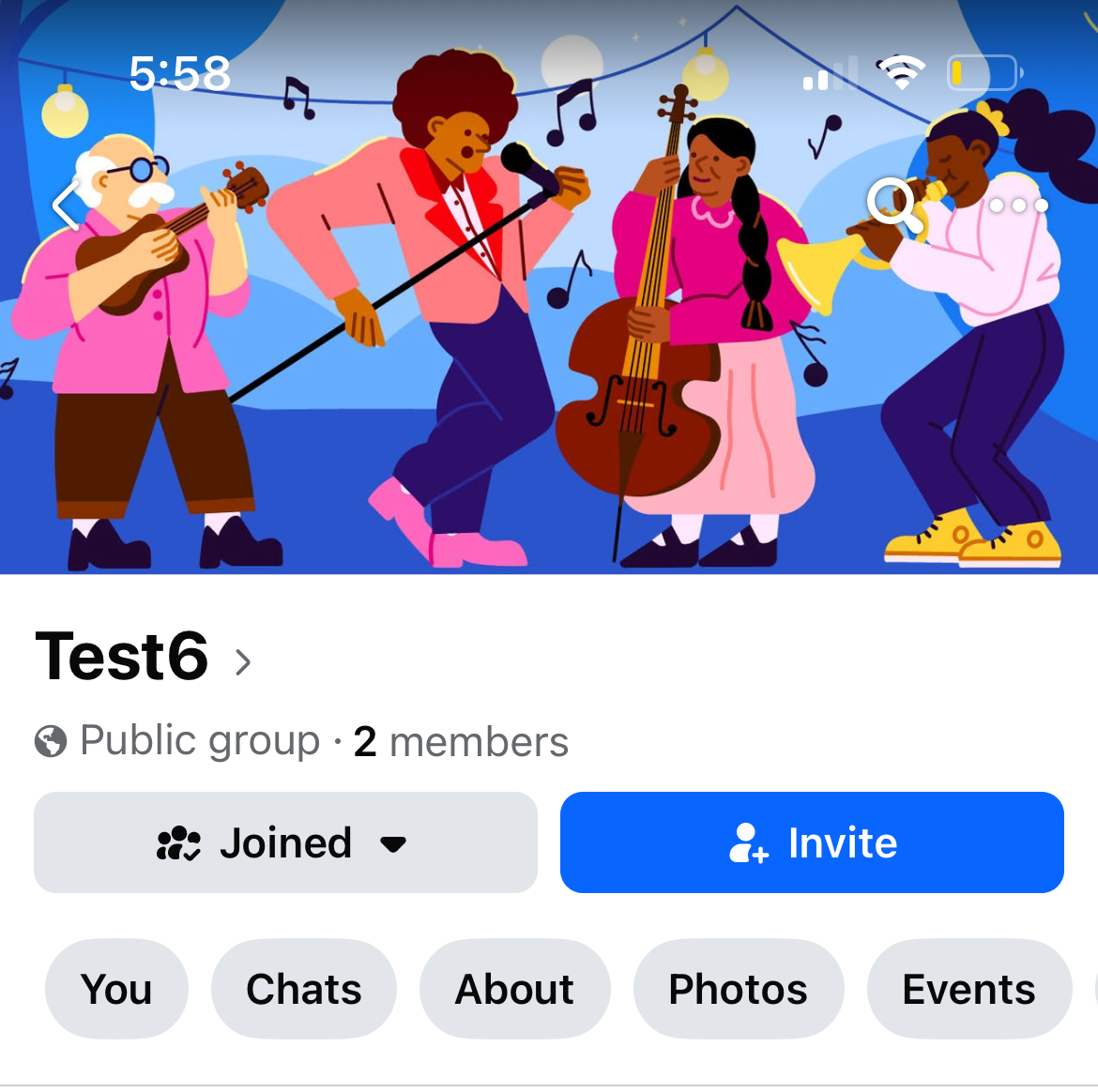}
    \label{fig:cover2}
  \end{subfigure}
  \caption{Representative Facebook Group cover-photo screenshots illustrating the three-dot ``Group tools'' menu icon in the upper-right corner.}
  \Description{Two side-by-side screenshots of two different Facebook Group cover-photos.
Left: group named 'Test1' (Public group, 2 members) with a gardening illustration; Joined and Invite buttons and tabs (You, Chats, About, Photos, Events) visible. A white three-dot menu icon sits at the upper-right over a light orange sky and has low contrast with the background.
Right: group 'Test6' (Public group, 2 members) with a band illustration; same buttons and tabs. The upper-right three-dot menu icon overlays a pale blue area and again appears low-contrast against the background.}
  \label{fig:group_meatball}
\end{figure}

Facebook Groups presented a similar visibility problem. On iOS, the horizontal \textit{menu} icon—which contains options like reporting—is rendered directly over the large decorative cover photo (Figure~\ref{fig:group_meatball}). Nicole, a participant with low-vision, attempted to flag a dummy group for spam but could not locate the icon; its low-contrast design blended into the photo and, as she put it, ``just disappeared.'' Our technical analysis\footnote{To assess potential low-contrast issues, we avoided relying on developer tools like Xcode's Accessibility Inspector, which cannot extract color values from graphical icons. Instead, we disabled True Tone and Night Shift on an iPhone 13, took lossless PNG screenshots, and used two separate color samplers to capture both the icon's foreground color and the surrounding background tones behind it. We used these values to test contrast sufficiency.} confirmed this was systemic: the icon failed the WCAG 2.1 non-text contrast threshold of 3:1 in five of Facebook's six\footnote{Link to the group cover photos we tested for contrast: \url{https://figshare.com/s/6b500311ac94d1b158e3}} default cover images. On X, the default menu icon on tweets also failed this test, but the platform's built-in ``Increase color contrast'' option in the accessibility setting rectified the issue\footnote{We provide screenshots for this here: https://figshare.com/s/4969cc7d674f2725d7a6}. In many settings, this 3:1 threshold is not merely a guideline: in the EU, it is a legal standard for public-sector apps under EN 301 549~\citep{en301549_2021} and is set to be extended to major private-sector platforms with a marketplace under the European Accessibility Act by 2025~\citep{EC_SWD_2022_410}.

\subsubsection{Workflow Navigability Challenges}
Once past the initial challenge of discovering the entry point for moderation, participants faced a second set of hurdles related to \emph{navigability}. Whereas \emph{discoverability} involved finding the correct icon or button to \emph{initiate} a moderation task, navigability here refers to completing the subsequent multi-step workflow. Challenges centered on \emph{unexpected focus order}, \emph{confusing interface cues}, and a \emph{lack of progress indicators}—together producing both learning and compliance costs.

\vspace{10pt}

\noindent\textbf{Unexpected Focus Order:} VoiceOver users depend on a consistent and predictable focus order when navigating mobile apps. Typically, when a new screen or dialog appears—such as after selecting a menu item—VoiceOver places the initial focus on a prominent element, often the ``Back'' button or the first actionable control. From this anchor point, users swipe right or left to move linearly through the interface. In our study, participants expected this behavior within moderation flows such as reporting a post or adjusting content settings.

\begin{figure}[htbp]
  \centering
  \begin{subfigure}[b]{0.47\linewidth}
    \centering
    \includegraphics[height=0.55\textheight,keepaspectratio]{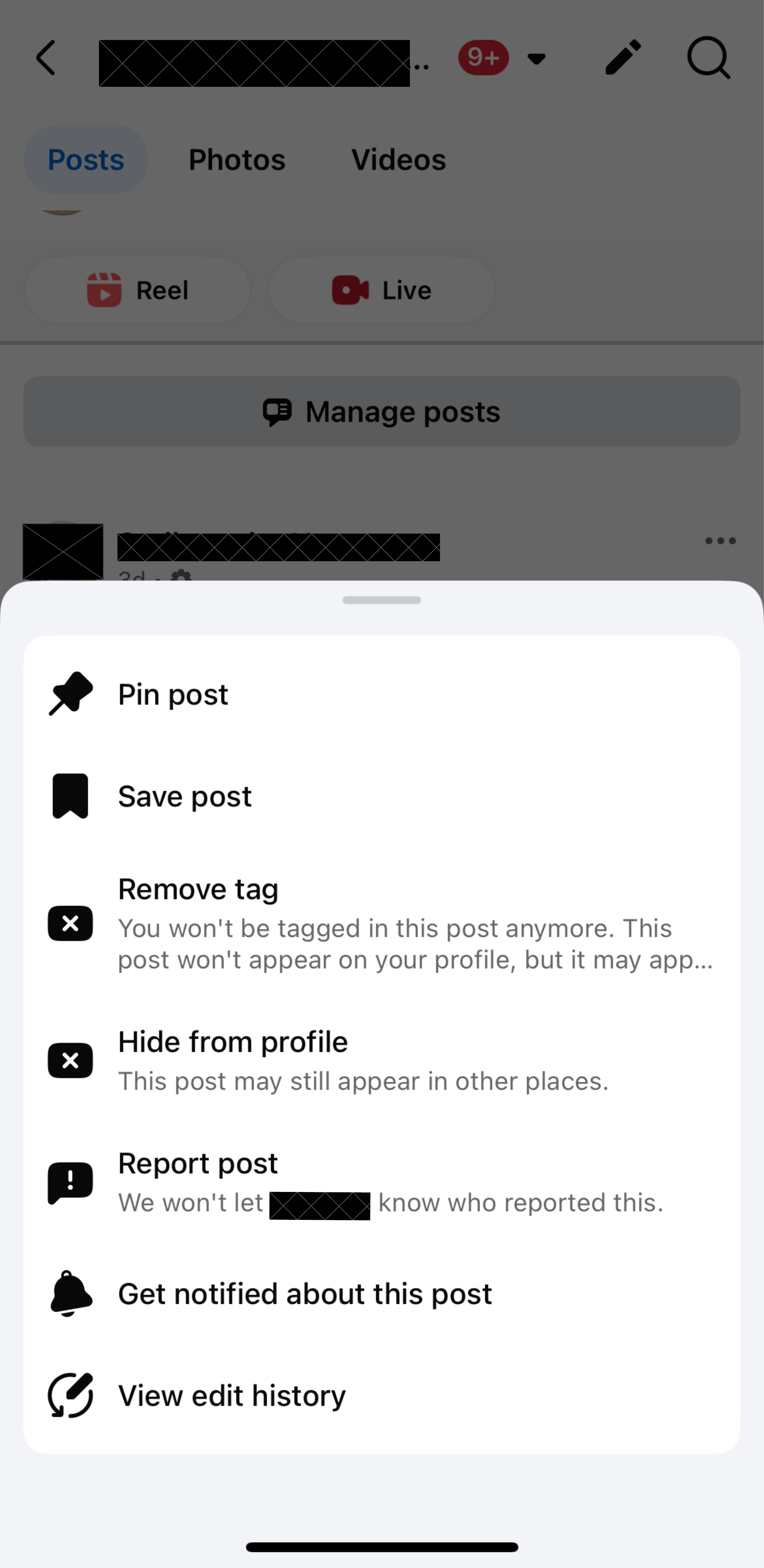}
    \caption{In-feed ``Report post'' action sheet that the user faces when they click on the ``More Options'' menu icon.}
    \label{fig:feed_report_popup}
  \end{subfigure}
  \hfill
  \begin{subfigure}[b]{0.47\linewidth}
    \centering
    \includegraphics[height=0.54\textheight,keepaspectratio]{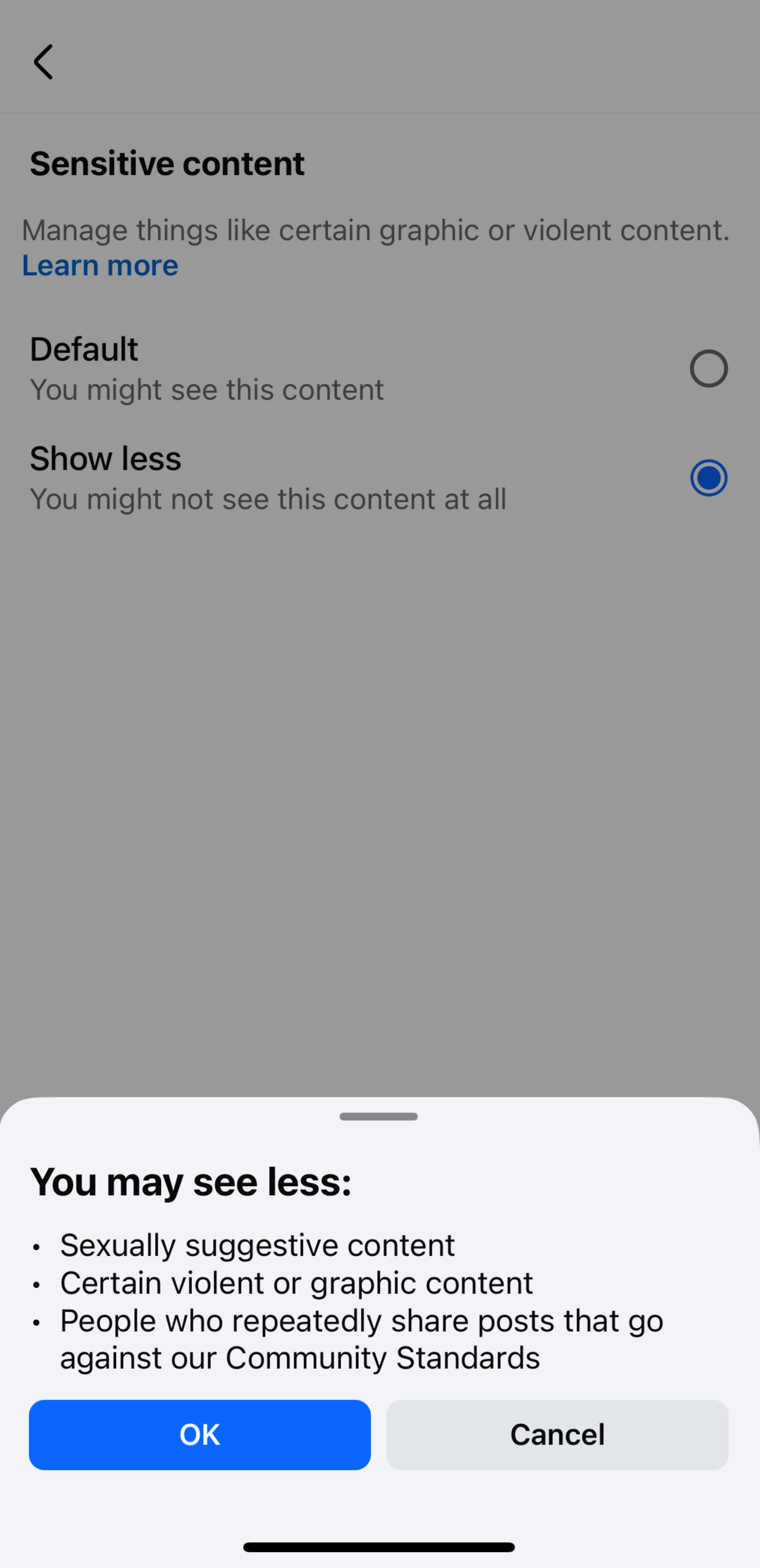}
    \caption{``Adult content'' preference panel that the user faces when trying to change the amount of recommended sensitive content.}
    \label{fig:content_pref_popup}
  \end{subfigure}

  \caption{Examples of pop-up panels that failed to auto-focus with VoiceOver, leaving users disoriented during key moderation and safety tasks.}
  \Description{Two side-by-side iOS screenshots of Facebook with dimmed overlays and bottom pop-up action controls.
  Left image: a half-screen bottom sheet sits over a darkened profile/post view. All controls are in the sheet, listing actions with icons—Pin post, Save post, Remove tag (with gray description), Hide from profile, Report post, Get notified about this post, and View edit history.
  Right image: the underlying 'Sensitive content' settings page is dimmed. A small bottom panel (about one-quarter screen high) contains a heading 'You may see less:' with a short bullet list and two buttons, OK and Cancel. All actionable controls are in this bottom panel.}
  \label{fig:popup_fb}
\end{figure}

These expectations often broke down in key moderation flows. We observed two types of unexpected focus order: (i) failure to focus on new views and (ii) illogical ordering of existing UI elements. An example of the former—after choosing to report a post on Facebook, a panel slid up from the bottom (Figure~\ref{fig:popup_fb}), yet VoiceOver failed to focus on it. Instead, VoiceOver focus floated unanchored, announcing only ``Dim background'' when users touched the screen's upper regions where content is typically expected. The actual controls were at the bottom, but without cues or automatic focus, participants were left disoriented.

Our participants struggled with this focus failure. 
Some mistook the ``Dim background'' overlay for a real control and kept tapping it.
Others realized it was inert but kept searching for the real control on the screen.  
A few became completely stuck and restarted the task. \emph{Of nine Facebook participants, eight} VoiceOver users encountered this problem at least once; only Nicole, who used a magnifier, did not. Confusion often led to self-blame before the design flaw became clear. As Samuel reflected after frustration: \emph{``I thought it was my mistake, but it’s really the design.''} These episodes show how a lapse in focus transfer cascades into disorientation, self-doubt, and repetition—an instance of the platform's \textit{psychological cost}.

\begin{figure}[ht]
  \centering
  \begin{subfigure}[b]{0.47\linewidth}
    \centering
    \includegraphics[height=0.55\textheight,keepaspectratio]{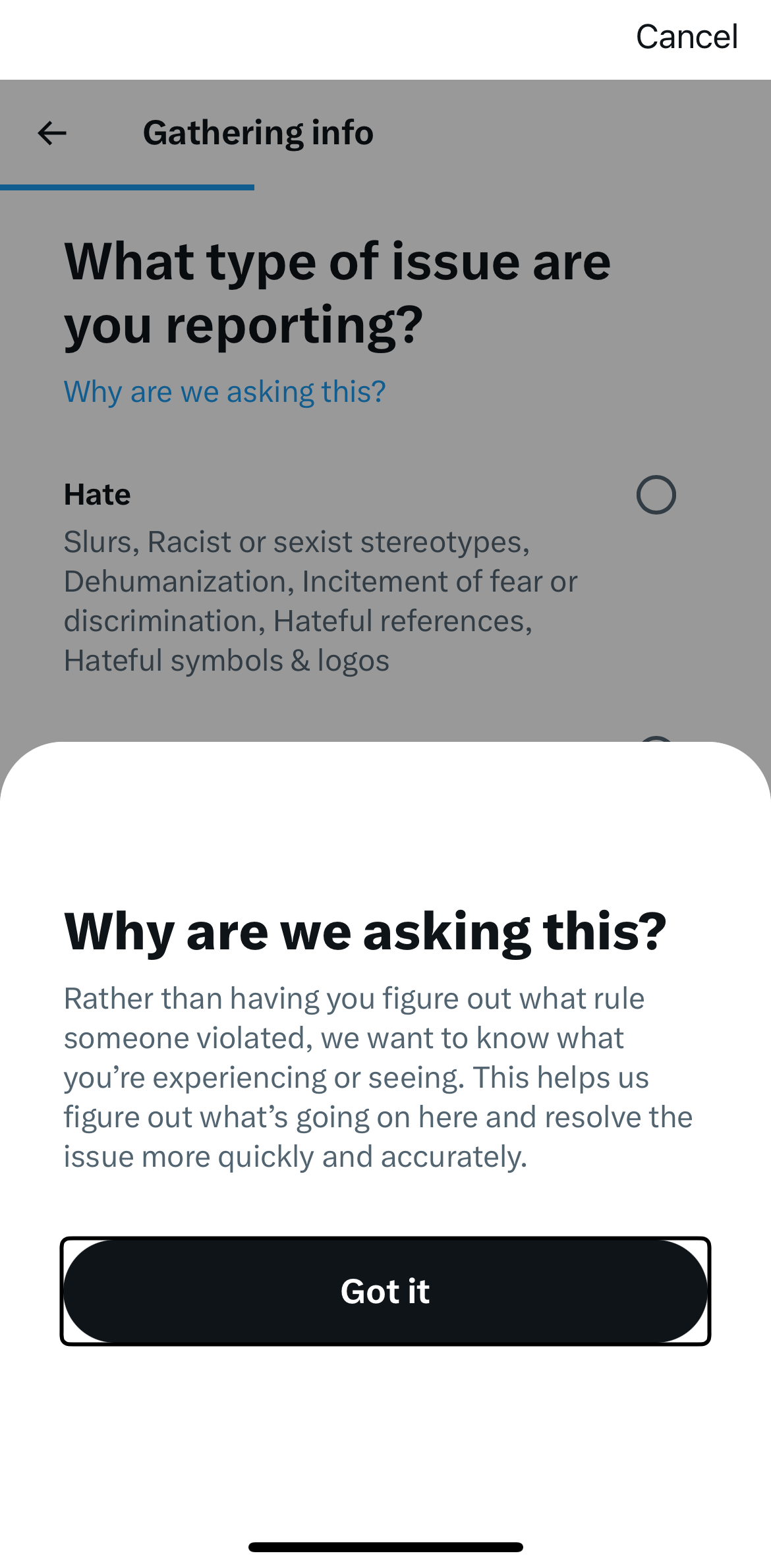}
    \label{fig:feed_report_popup}
  \end{subfigure}
  \hfill
  \begin{subfigure}[b]{0.47\linewidth}
    \centering
    \includegraphics[height=0.55\textheight,keepaspectratio]{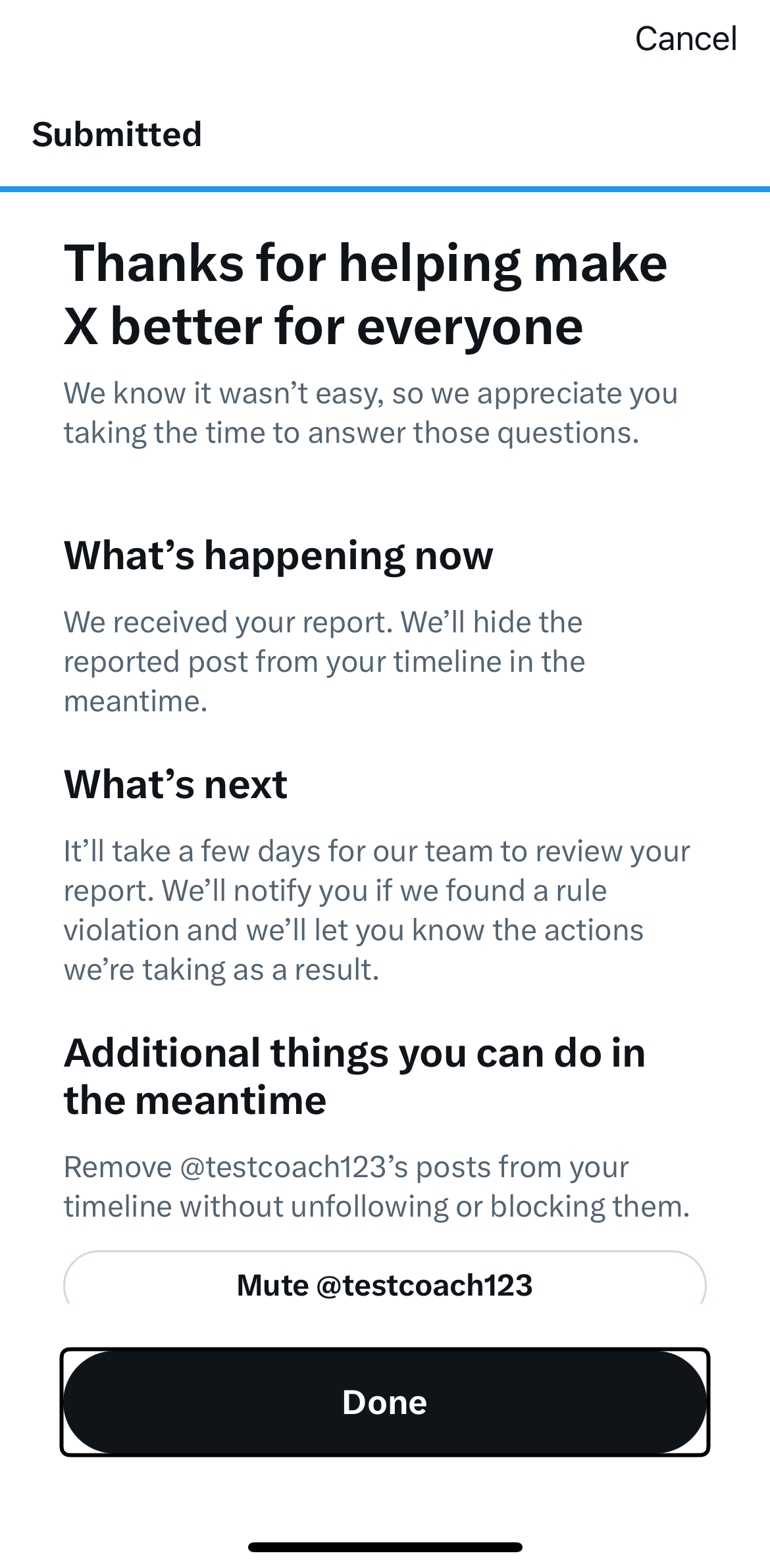}
    \label{fig:X_popup}
  \end{subfigure}

  \caption{X's pop-up panels: VoiceOver focusing directly on primary controls in additional-information wizard (left) and in a post-submission confirmation page (right).}
  \Description{Two iOS screenshots from X's reporting flow.
  Left image: the main screen is dimmed by a dark overlay while a rounded bottom sheet covers about half the screen. The sheet contains a short explanation and a single large primary button at the bottom on which VoiceOver focus is placed; the header shows Cancel option.
  Right image: a submission confirmation page (no overlay) with brief sections of text and two controls near the bottom—an outline action to mute the account and a prominent Done button on which VoiceOver focus is placed.}
  \label{fig:popup_twitter}
\end{figure}

Focus and navigation problems extended beyond this initial hurdle. Once participants succeeded in latching onto the pop-up panel, the focus engine still betrayed a sight-first logic, especially on X\footnote{We document the entire workflow for reporting a dummy post using VoiceOver here: https://figshare.com/s/20940fa4b9f7d51c3ad9}: it frequently skipped the first actionable element and jumped straight to what a designer might consider the ``primary'' control. The reporting wizard begins with a prompt—\emph{``Which type of issue are you reporting?''}—followed by a short help link labeled \emph{``Why are we asking this?''}, as shown in Figure~\ref{fig:popup_twitter}. When Cristian clicked on that link to learn more, VoiceOver did not land at the explanatory text; instead it leapt past the paragraph and settled on the final \emph{``Got it''} button. Cristian expressed: \textit{``Wait, I asked for the explanation and it’s already telling me `Got it'?...I never heard the explanation!’’}

A similar focus shortcut appeared at the \emph{end} of X's reporting flow. After Mateo submitted a hate speech report, the confirmation screen offered follow-up actions—mute, block, or limit notifications—stacked above an unmistakable \emph{``Done''} button. VoiceOver, however, landed directly on \emph{Done}. Unaware of the additional safeguards, Mateo simply double-tapped and exited:\textit{``It said `Done button,' so I pressed it...So you're telling me there's more options further up the page?''}

\noindent
Focus oddities also amplified the cost of mistakes. Reporting wizards usually require several categorical choices across different interface pages; if a user realized on page 2 that they had picked the wrong category on page 1, there was often no reliable way back. X's workflow, for instance, omits a \emph{``Back''} link after the first step—Grant discovered this the hard way and had to abandon the wizard entirely: \textit{``I tapped the wrong reason and...[tried to go back but] found no back button...had to start from scratch.''}

Even escaping the reporting wizard, once initiated, was found to be tricky on X. Many participants sought a \emph{``Cancel''} or \emph{``Close''} control in the upper-left corner—mirroring standard iOS convention—only to find the control tucked away at the upper-right on X but announced \emph{last} in the swipe order. 

Collectively, these focus and navigation hiccups turn minor deviations from platform conventions into a measurable burden: they create \emph{learning costs} by obscuring where focus should land and masking entry-point/help text, and they increase \emph{compliance costs} by forcing extra swipes, detours, and restarts—often breaking the user's momentum.

\vspace{10pt}

\noindent\textbf{Redundant Cues and Hidden Confirmations:} X’s reporting wizard presented a different challenge: excessive verbosity coupled with a lack of audible confirmation for selected options. When navigating the category choices while reporting a post, VoiceOver exhibited a redundant pattern: it first read the category heading, then its descriptive sentence, and finally repeated \emph{both} the heading and description together, appending ``radio button.'' The layout of the reporting wizard created three problems:

First, the wizard required users to swipe through long, repetitive blocks of text to reach the ``Next'' button—a slog Cristian likened to \textit{``wading through every block before you reach `Next.' ''} Because the section headings could not be skipped by VoiceOver without going into the description, participants could not jump by headings and thus had no quick way to skip ahead.  

Second, compounding this slowness was a critical lack of auditory feedback. When a participant double-tapped to select an option, the interface accepted the input but offered no audible acknowledgement like ``checked.'' Mateo explained the resulting doubt: \textit{``I tapped while it was still explaining, and it never said `checked'...so I had no idea it took my answer. You can see it though right?''}.

Finally, the redundant structure of the announcement itself created confusion. After hearing the same content spoken twice, participants sometimes mistook the repetition for a new, distinct category. Cristian echoed this: \textit{``It read the title, the paragraph, then both together. The second time it read it I thought it was a different category...I didn't pay close attention to it until I went back to the description again.''} This combination of forced verbosity, missing confirmation, and confusing repetition charged a significant \textit{compliance cost} in the form of cognitive burden.

\vspace{10pt}

\noindent\textbf{Uncertain Progress:} Because neither platform clearly states how many screens lie ahead, first-time reporters often mistook mid-flow pages for the end of the process.  Frank who performed the task on Facebook admitted, \textit{``If I'm a first-time user...I would just find `Close' and not realize they're asking me [another question].''}  Adriana echoed this sentiment, saying she would prefer cues like \textit{``Step 1 of 3''} upfront so she could gauge whether a post was \textit{``worth the effort of reporting.''}  X does show a progress bar, but as a percentage; some participants failed to map ``33\%'' to ``one-third done.''  Grant thought the concept was ``okay'' yet suggested that question or page counts—``1 of 3''—would be clearer.  Without such cues, participants grew weary; Adriana confessed on the third page of reporting the subcategory, \textit{``I’m annoyed...I will put my phone down [at this point] and leave it alone.''}  

\vspace{5pt}
\noindent Despite the initial fatigue, persevering participants highlighted several positive design elements that tempered their frustration. The workflows themselves, once located, were sometimes praised for their clarity. Vasu called the flow \textit{``smooth'' and ``comprehensive''} after finding it; the granularity of options conveyed a sense of legitimacy for him. Jessica appreciated that Facebook's wizard \textit{``asked questions to make sure I gathered all the correct information,''} which suggested reports were taken seriously.

Crucially, clear feedback and confirmation were highly valued. Positional cues like ``1 of 5'' on X's list option items anchored users and reduced guesswork, while summarizing reporting details just before submission and explicit spoken receipts about receiving the report—like Facebook’s ``Thanks for reporting this post. Next step..awaiting review...''—assured users like Frank that their complaint was logged. Similarly, a simple closing line on X, ``Thanks for helping make X better for everyone,'' provided a welcome sense of closure for Grant. Finally, participants applauded the ease of access to post moderation controls, such as X's integrated options to immediately \textit{``Mute the person...or block them,''} (Figure \ref{fig:block_hidden}, right) which Maya saw as a considerate touch that restored user control. These positive examples show that clear cues, granular options, and empowering follow-up actions can significantly improve the user experience and foster a sense that one's effort can yield tangible results.

\subsection{Uncertainty in User Controls}
\label{sec:findings:inconsistencies}
Beyond challenges in discovering and navigating specific workflows, participants encountered numerous \emph{design uncertainties} across platforms that created significant \emph{cognitive friction}. Some key issues were \emph{inconsistent labeling}, \emph{lack of preview information}, and \emph{illogical placement and inaccessibility of controls}.

Participants experienced this friction from the moment they attempted to open options menus. A recurring theme was the inconsistent labeling of the overflow menu icon across different contexts on Facebook. While the same 
``\raisebox{0.5ex}{\meatballicon[0.4ex]{lightgray}}''
icon is used universally for post, profile, and group options, VoiceOver announced a different label in each case – i.e., ``More options'' on a post, ``See More'' on a profile, and ``Group tools'' on a group page. This lack of uniformity caused confusion and extra mental effort. Laverne captured the mental tax this imposes: \begin{quote}
\textit{``[For profile] It says `see more.' And the other one [post] was `more options'... And for another... ‘group’ something...So with three completely different things [for the same icon], you would have to get comfortable and remember at my age.''}
\end{quote}
Several participants found this especially frustrating because it imposed more uncertainty for users with BLV compared to sighted users. Vasu defended the context-specific labels at first, but ultimately conceded that a similarly sounding descriptive label would be far better for everyone: \begin{quote}
\textit{``If they call it `Tools' or `More options' everywhere, that would have been the best... When you [the interviewer who is sighted] see it you know immediately [that it is an option for menu items]. It should be the same for us.''}
\end{quote} 
Other participants noted that even when they found the icon, the surrounding context did little to signal its purpose as a menu button. As Adriana explained after swiping past the menu icon without recognizing it,
\textit{```See more' doesn’t tell me anything. It just says `See more'\dots{} See more of what?''}

When an action as fundamental as opening an options menu is obscured by inconsistent or opaque labeling, users with BLV must either rely on trial-and-error exploration or abandon the feature altogether. Such friction not only slows task completion but also reinforces for users with visual impairments a sense that the interface was not designed with non-visual navigation in mind.

Once participants finally reached the moderation options, they still had to gamble on what each feature (like flagging, blocking) would actually do. Neither platform previews the consequences: X surfaces an explanation only after the button is pressed, while Facebook reveals the details only once the action is complete. In that vacuum, everyone based their understanding of moderation features, often incorrectly, on their mental models of similar tools they had encountered in other contexts. For example, Laverne (63, who mostly relies on iMessage and phone calls) tried to unblock a user on Facebook by hunting for the person’s profile\footnote{Unblocking a person on Facebook is currently only possible by finding the exact option in the privacy settings.}—mirroring how she would unblock a contact on her phone—because, as she reasoned, \begin{quote}\textit{``I thought it’d be like blocking a number: if I block someone they can’t text me...[but] I can still message them. I have no idea how blocking [on Facebook] works.''}\end{quote} Jessica carried a different assumption: \begin{quote}\textit{``I didn’t know you could block someone who isn’t your friend... I’d never have learned that if we hadn’t done this task.''}\end{quote} Lacking immediate, task-time descriptions, participants built inconsistent mental models of \emph{Block}, \emph{Mute}, and \emph{Report}, discovering the truth only through trial and error—a design gap that left every user guessing about critical safety controls.

\begin{figure}[ht]
  \centering
  \begin{subfigure}[b]{0.47\linewidth}
    \centering
    \includegraphics[height=0.55\textheight,keepaspectratio]{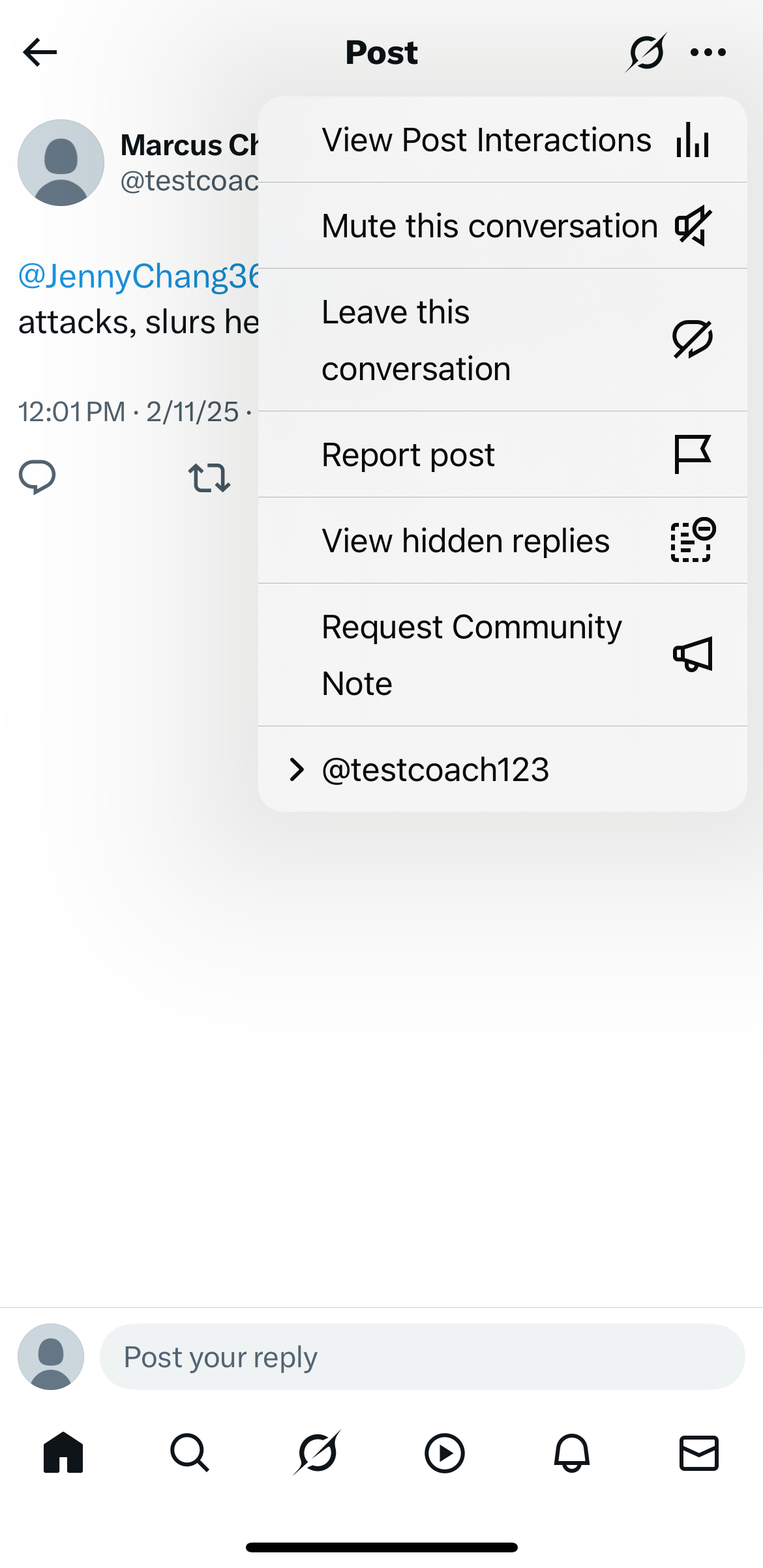}
    \label{fig:hidden1}
  \end{subfigure}
  \hfill
  \begin{subfigure}[b]{0.47\linewidth}
    \centering
    \includegraphics[height=0.55\textheight,keepaspectratio]{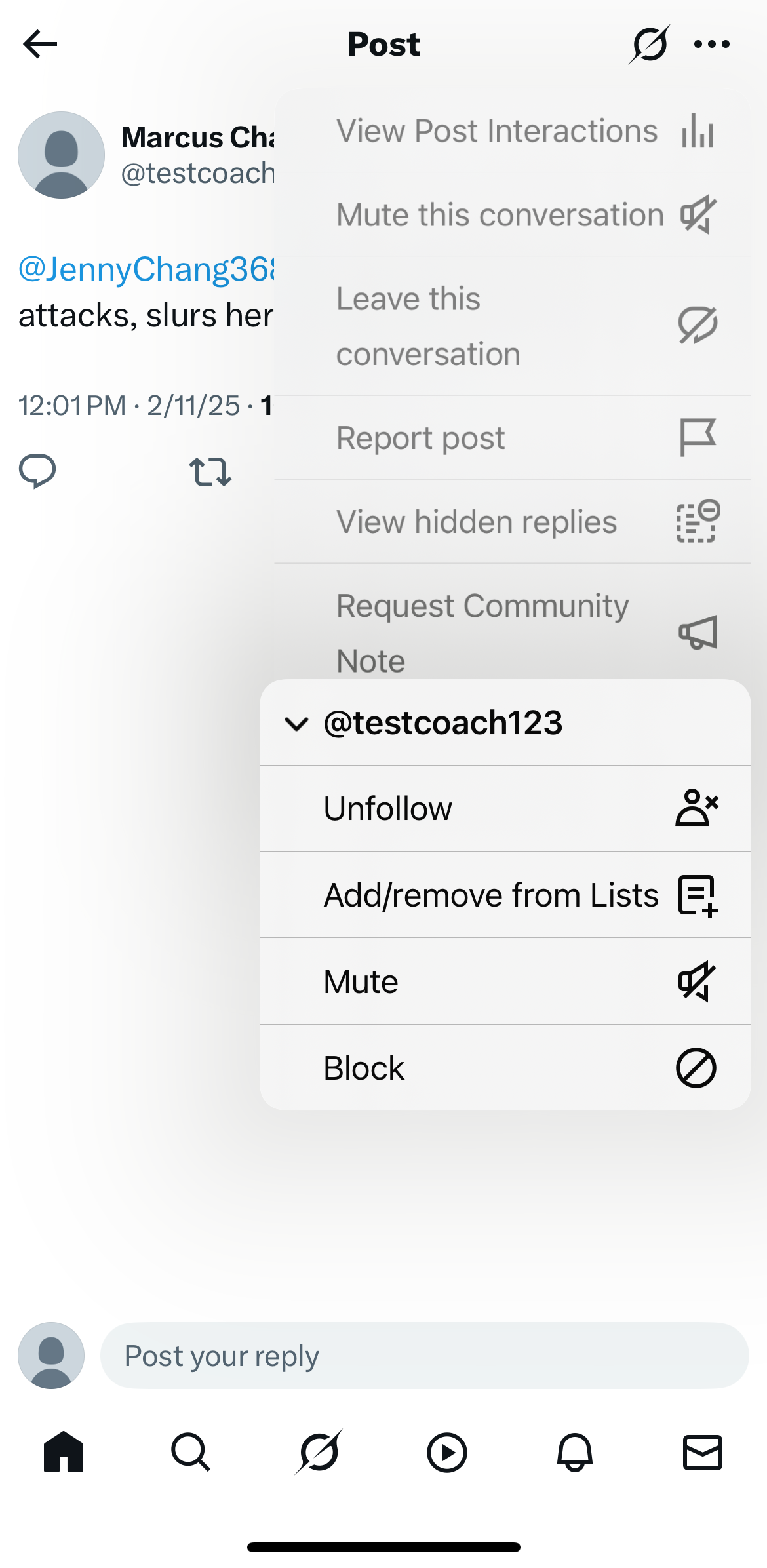}
    \label{fig:hidden2}
  \end{subfigure}

  \caption{On X, the ``Block'' feature is two layers deep: users must first open the post’s \textit{More actions} menu (left) and then expand the ambiguous @username submenu (right) before the option appears—hiding a critical safety control along a path that offers little contextual cue for VoiceOver users.}
  \Description{Two iOS screenshots of an X post with menus options expanded over a dimmed background which has the post.
  Left: the menu options are dropped down in front of the post as a tall sheet; a collapsed row with the account handle appears at the bottom and no Block option is visible.
  Right: after expanding that handle, which is a drowdown, a second submenu appears showing 'Unfollow', 'Add/remove from Lists', 'Mute', and 'Block' (Block is only in this second layer)}
  \label{fig:block_hidden}
\end{figure}

Most participants' first instinct when asked to ``block the (fictional) hateful user'' was to look for the control right on the offending post itself: \emph{eleven of fifteen} swiped through the tweet or Facebook post, opened the ``More options'' menu, and were surprised when no \textit{`Block'} command appeared. ``\textit{I am seeing if there's [a possibility to] block user from the post...I thought they would have it here},'' Patricia complained after swiping through all the options provided on the menu options in the (fictional) hateful post before realizing she had to detour to the author's profile. X technically offers the affordance many wanted—a block button nested in the actions attached to every tweet—but discoverability was uneven. One participant, Mateo, celebrated it as exactly what Facebook lacks: ``\textit{I love that they put the block right there...somebody got you so aggravated...block right there}.'' Yet others missed the same shortcut because it sits inside a secondary submenu inside the ``More actions'' menu (see Figure~\ref{fig:block_hidden}). Shreya's first visual sweep through the tweet options produced only silence—``\textit{Oh, there’s no option for block from the more [actions] on the post itself}''—and Cristian concluded the feature was ``\textit{a little hidden when it could be under more actions [on the post itself]}.''

 Even more elusive than post-level shortcuts were the platform-wide settings themselves. During tasks requiring participants to \emph{unblock a previously blocked user} or to \emph{adjust sensitive content filters}, \textbf{nine of fifteen} participants failed to locate the controls, as they were hidden deep within sprawling category trees. Jamila quipped that finding them felt like \textit{``hunting for an Easter egg.''} The in-app search function which many turned to as a last resort, offered little help by failing to interpret user intent. For example, on Facebook, participants attempting the unblocking task searched for ``unblock'' in the privacy settings search bar, which yielded no results. The correct term, ``block,'' was required to find the list of blocked users. Adriana called this literal search function ``stupid,'' as it could not infer her goal. 
 
A parallel failure played out on X. When tasked with changing their settings to see less sensitive material, participants entered logical search terms like ``Adult content,'' ``Sexually explicit,'' and even ``Sensitive content.'' Yet the search failed to return the relevant toggle, which was filed under the precise label ``Display media that may contain \textit{sensitive content}.'' These rigid and ineffective search mechanisms made platform moderation affordances feel not just hidden, but functionally unavailable to our participants.

Taken together, these uncertainties translate into a concrete burden: they create \emph{learning costs}—users must infer meanings for identically-shaped icons across contexts, act without task-time previews, and wrestle with literal, intent-blind search—and they raise \emph{compliance costs} by forcing detours (e.g., from posts to profiles), multi-level submenu traversal, and repeated attempts (e.g., 11/15 first looked for \textit{Block} on the post; 9/15 could not locate the relevant settings). This led to guessing, backtracking, and often abandonment of the very safety tools meant to empower users.

\subsection{Adaptation, Emotional Labor, and Perceived Agency} 
\label{sec:findings:labor_agency}
We next turn from pragmatic and functional barriers to users' elucidation of their problems. We document the adaptations participants employed, the extra interaction work these entailed, and the emotional consequences that followed. These constitute the psychological dimension of administrative burden in end-user safety tasks.

\subsubsection{User Workarounds}
Participants described inventive workarounds to navigate content moderation interfaces when standard methods fell short. When users on X swiped past the menu option that reveals more controls (including reporting), they resorted to touch exploration—sliding a finger across the screen surface to identify and interact with elements by vocal feedback. 

\begin{figure}[ht]
  \centering
  \begin{subfigure}[b]{0.45\linewidth}
    \centering
    \includegraphics[height=0.5\textheight,keepaspectratio]{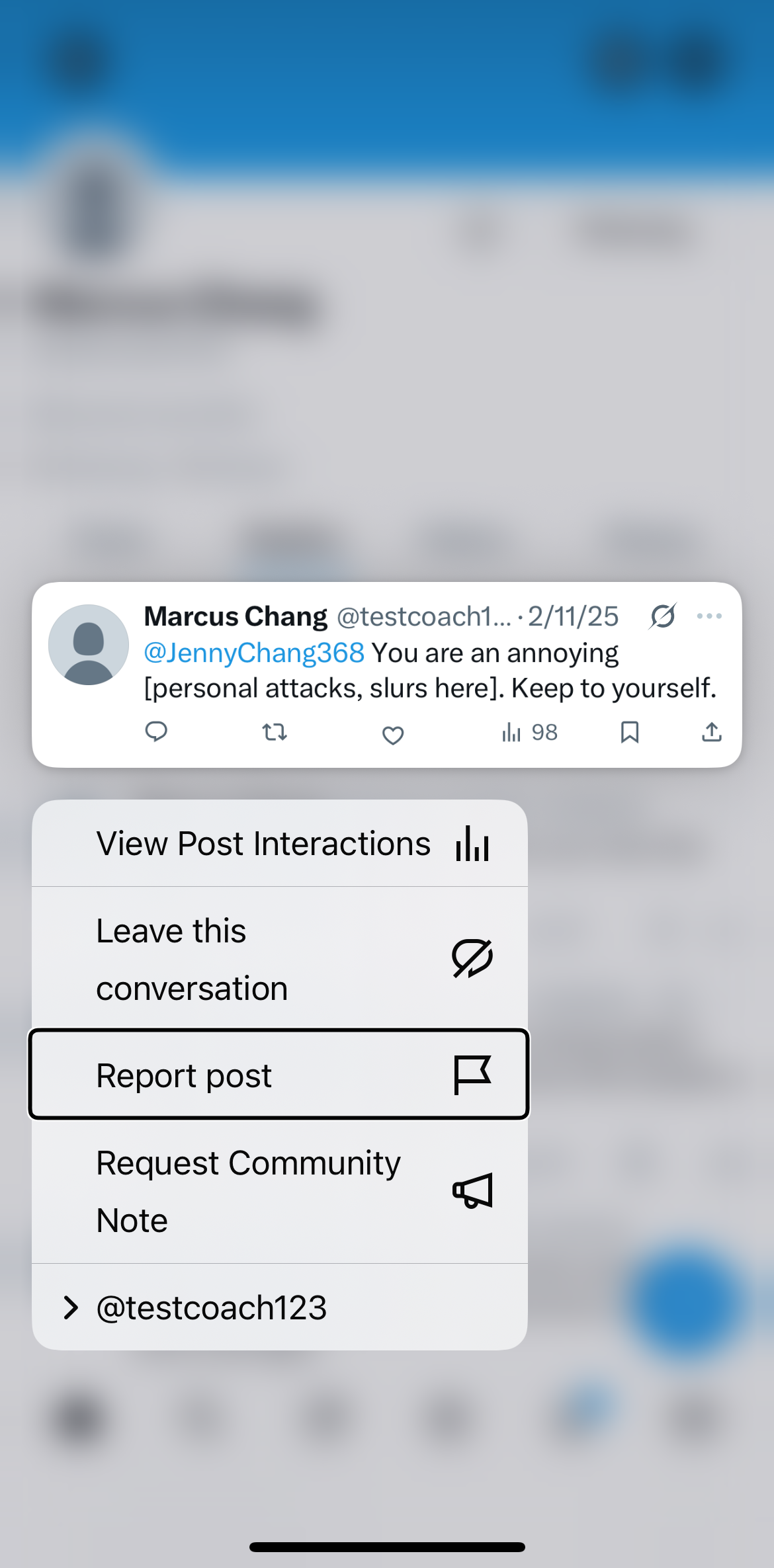}
    \label{fig:feed_report_popup_carlos}
    
  \end{subfigure}
  \hfill
  \begin{subfigure}[b]{0.47\linewidth}
    \centering
    \includegraphics[height=0.38\textheight]{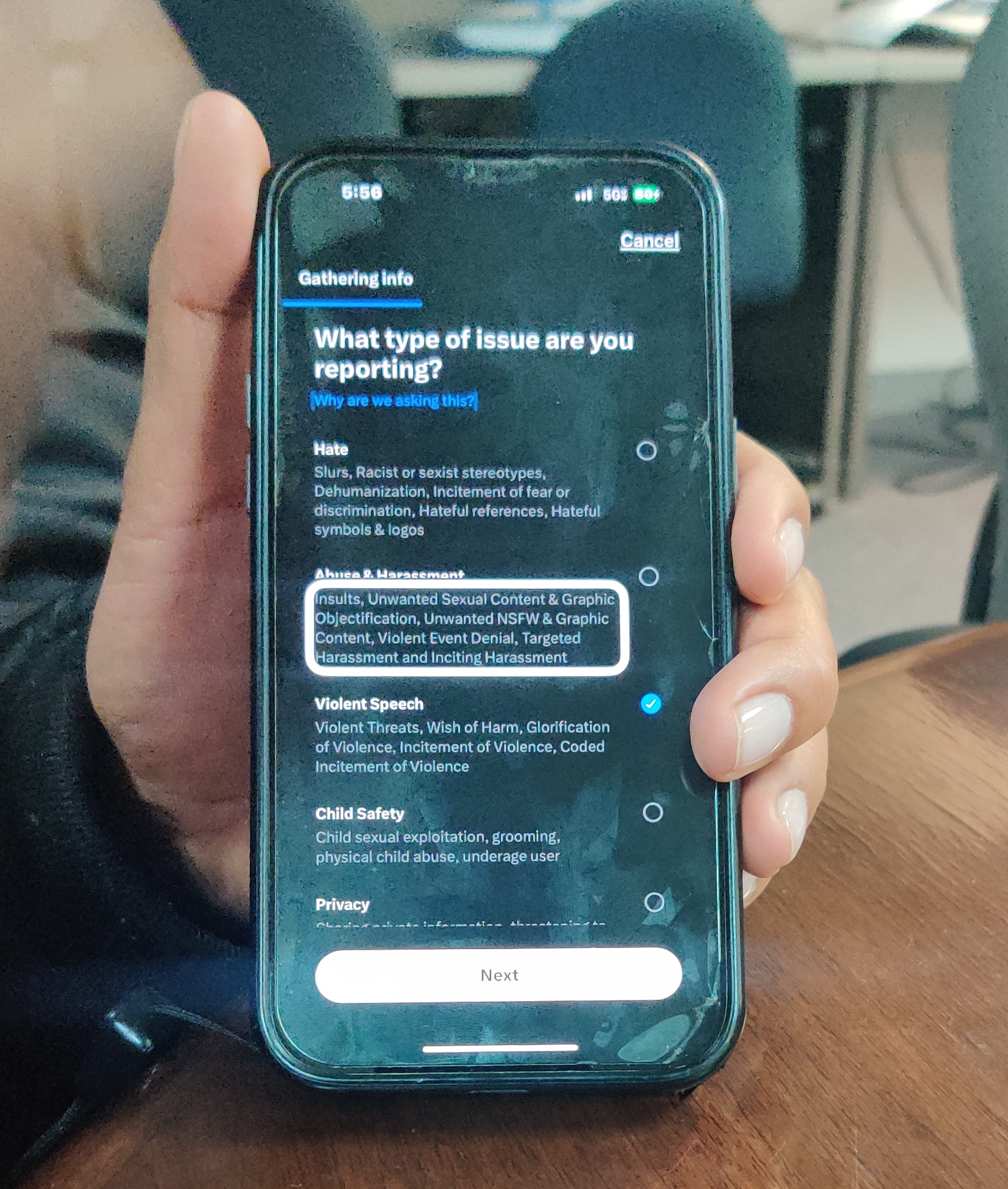}
    \label{fig:X_popup_carlos}
  \end{subfigure}

  \caption{Workarounds adopted by Cristian while using X: Double-tap and hold brings options available haptic actions (left) and using his middle finger as a marker for the current action line (right).}
  \Description{Two images are visible.
  Left: iOS screenshot of X with the timeline dimmed and a post actions menu open; the 'Report post' row is highlighted among other options.
  Right: photo of a person holding an iPhone on X's reporting screen (dark mode). Their thumb rests on the display as a physical marker to remember where VoiceOver's current focus is; a 'Next' button is visible at the bottom.}
  \label{fig:carlos_adaptation_twitter}
\end{figure}

Cristian—a highly proficient VoiceOver user and a trainer for using assistive technology—bypassed the elusive menu icon on X by placing his finger on a post and performing a double-tap-and-hold gesture (Figure~\ref{fig:carlos_adaptation_twitter}, left) (equivalent to a long press) to instantly reveal a context menu with options including ``Report Post,'' a workaround he learned by himself to avoid the lengthy sequential swipes needed otherwise. Because such non-obvious shortcuts are easily missed by novices who rely solely on flicking through content, Cristian now teaches this technique to other blind users, framing it as a vital strategy for uncovering hidden actions within X’s interface.

Low-vision participants demonstrated their own arsenal of adaptation strategies. Some preferred magnifying content visually over using screen readers. For instance, Shreya (who has extremely high myopia) relied on iOS’s Display Zoom and large text settings to enlarge posts, only toggling VoiceOver on for certain tasks. She expressed a personal attachment to ``seeing'' content, trying to remain a visual user despite her deteriorating sight. Nicole likewise used a physical 11x magnifying glass with her phone, holding it over the screen to read fine print or interface labels. This workaround helped her avoid the layout disruptions caused by oversized text settings. \textit{``I have to use my magnifier all the time on my phone...it's just habit,''} Nicole admitted, acknowledging that increasing the on-screen font to a comfortable size often made content run off the screen. For her, a handheld magnifier provided a more predictable view of the screen layout that would otherwise be cut off or require panning.

When social media apps proved too confusing even with previously discussed workarounds, participants turned to virtual assistance tools like `Be My Eyes'\footnote{https://www.bemyeyes.com} (which connects to a volunteer or AI who can describe the screen) or Siri. Cristian mentioned that if he ever truly got stuck, he would take a screenshot of the mobile app layout and ask the `Be my AI' feature of the `Be my Eyes' app to describe the screen for him. He also described a personal technique of using a second finger as a physical ``marker'' on the screen to help track his current position while navigating or checking other information (like notifications or time) (Figure~\ref{fig:carlos_adaptation_twitter}, right). Cristian suggested this highlighted the need for built-in features, like a VoiceOver ``hot spot'' option, to temporarily bookmark a location on the screen and easily return to it after interruptions, reducing the mental labor of re-orienting.

Through techniques like touch exploration, memorizing screen layouts, utilizing zoom and magnification, or seeking human assistance, BLV users exhibited resilience and ingenuity. These adaptations empowered participants to complete reporting and blocking tasks. However, employing these workarounds often required extra effort, expertise, and patience—resources that not all users have in equal measure. This personal labor compensated for design shortcomings, allowing our participants to exert agency over content moderation actions \textit{in spite of} the platforms, rather than because of them. 

\subsubsection{Interaction Overload and Participant Fatigue}

Even with clever strategies, the interaction burden of content moderation tasks was strikingly high. Participants frequently encountered workflows that were labyrinthine, requiring dozens of gestures and trial-and-error attempts to accomplish what seemed like a simple goal. For instance, changing a privacy setting or unblocking a previously blocked user could demand navigating through multiple layers of menus, each involving sequential swipes, double-taps, and back-tracking through screens. We observed that without using the search function to navigate directly to a setting, users might have to perform in the order of \emph{thirty} or more discrete actions to reach a deeply nested option, assuming that they don't take detours, which would further exacerbate tasks. 

Our participants voiced exhaustion at these drawn-out procedures. Several described feeling as if they were ``walking in circles'' through menu trees and when they found it, they `didn't even have an idea how they got there'. One user quipped that she had to attempt the task three separate times initially before learning the correct path to her blocked list. During a Facebook task to unblock a user, Samuel became so frustrated by the maze of options that he eventually asked Siri for help, treating the problem as one big question to be answered externally. He confessed that without an external shortcut, he felt lost in Facebook’s interface: \textit{``I don’t have knowledge of the whole interface,''} he said, after combing through Settings fruitlessly. The number of sub-menu categories and lack of clear pointers wore him down. Describing the convoluted menu system, Samuel joked— \textit{``...there are so many options, and more options inside them. They should [instead] call it `Maze-book' or something.”}

\subsubsection{Emotional labor and Perceived Lack of Agency}
Beyond the practical challenges of interacting with moderation tools described in the previous sections, participants’ experiences with reporting and moderation carried an emotional labor. Engaging with these tools – often in the context of hateful or abusive content – sometimes provoked anxiety, distress, and a diminished sense of agency. Several users expressed trepidation about even initiating a report. Laverne, for example, was so wary of ``doing something wrong'' on Facebook that she hesitated to try any moderation settings at all on her own profile. She had heard rumors about users being penalized for acting in suspicious ways, which made her afraid to experiment or make a mistake. As she put it: \begin{quote}
\textit{``I’m afraid to just touch anything..I don’t know what I’m doing. I've heard people say you can get put in Facebook prison.''}
\end{quote}
Laverne shared that she was on Facebook only to read posts from her granddaughter and was apprehensive about losing access to the platform. 

Shreya and Jamila, both of whom had confronted harassment online, described feeling disheartened and even mocked by the outcome of reporting incidents. They had dutifully reported abusive posts directed at them, only to see no visible action taken against the perpetrators. Jamila recounted reporting several hate-filled comments on her post in a local community group, but, ``usually nothing comes of it,'' she said–the toxic posts remained up, and moderators or Facebook staff often responded with silence or a generic rationale about free speech.
These experiences echo prior findings on post-reporting grievances of sighted users~\citep{zhang2023cleaning}, but they emphasize their salience for this especially vulnerable population as well.

Moments like this imbued a sense of futility. Participants felt that they were doing their part by courageously flagging abuse, yet the platforms did not reciprocate with adequate support or resolution. Shreya shared that seeing rampant harassment go unchecked on X made her withdraw from interacting altogether–\textit{``why put myself at risk of further harassment if it [X] would not back me up?,''} she reasoned. 

This emotional disillusionment eroded the trust that reporting mechanisms would actually help; instead, using them sometimes felt like reopening wounds for no benefit according to some participants. In addition to fear and disappointment, many participants voiced frustration at the false sense of agency created by these tools. Social media platforms outwardly empower users to classify and curb harmful content, but in practice our participants often felt like the burden of moderation was shifted onto them without real power. Participants bristled at being forced to categorize the offense themselves. After navigating a reporting form on Facebook, Vasu questioned why the onus was on him to decide if something was ``harassment'' or ``hate speech,'' asking rhetorically, \textit{``Why do I have to do it? Isn’t it their job?''} Patricia felt that forcing a user to select the precise category of abuse was asking them to \textit{``endure [the pain]...for your [platform’s] convenience.''} In her eyes, the reporting process was designed for the platform's convenience rather than the user's well being. These reactions mirror prior work showing that flagging/reporting systems conscript end-users and volunteer moderators into governance roles while platforms provide limited follow-through or support—leaving people to shoulder moderation labor with little feedback or remedy~\citep{Crawford2016Flag,Matias2019Civic,Jhaver2019removal}.

Encounters with illusory controls—such as X's \emph{Report community} link that merely redirects\footnote{A demonstration of this redirection is provided here: https://figshare.com/s/4be546a24b0610accde7} to a help article—further eroded participants’ faith that moderation tools would actually address their concerns. Several participants suspected their submissions served mainly to refine platform's content curation algorithms rather than to secure personal relief. In sum, we documented our participants' emotional cost of this ambiguous labor: anxiety over potential backlash, discouragement when no feedback arrived, and anger at having to police abuse without meaningful support.

%% file: 06-discussion-v2.tex
\section{Discussion}
\label{sec:discussion}

We discuss our results through two intertwined manifestations of administrative burden in end-user safety tasks for participants: \emph{uncertainty} and \emph{labor}. By uncertainty we mean lack of clarity in situations where participants have to figure out whether safety controls exist, where they are, and what they will do, which disrupts mental models and progress. By labor we mean the effort users expend to make sense of a situation and reduce uncertainty so they can proceed—repeating gestures, taking detours or restarts, and devising workarounds or seeking help. We foreground these aspects because they recur across our data and let us trace how administrative burden's \emph{learning, compliance and psychological} costs, detailed in section \ref{sec:findings}, interleave in practice. The subsections that follow explicate where these costs emerge, intertwine, and cascade across steps, and we provide suggestions to mitigate these costs.

\subsection{Uncertainty: Learnability, Opacity, and Mental-Model Drift}

We begin with \emph{uncertainty} because, in our data, not knowing what a safety control does, where it lives, or how the control functions persistently incurred \emph{learning costs} at the outset of safety work. These learning costs then spilled into \emph{compliance} and \emph{psychological} costs as people probed, backtracked, and second-guessed their actions. In our analysis, uncertainty surfaced at five recurrent points that we detail below: (1) whether a control existed at all; (2) inconsistent names applied to the same icon; (3) missing cues about how the tool will function before commitment; (4) vague or misleading progress cues; and (5) opaque, hard-to-trace outcomes. Together these form end-to-end transparency gaps—a mix of \emph{semantic opacity}—where inconsistent labels and intent-blind search undercut discoverability, and \emph{procedural opacity}—where users get little insight into what a control will do and only vague signals of progress, completion, or outcome. These gaps disrupt users' mental models and transform safety work to precarious guesswork.

\textbf{(1) Existence uncertainty.} Participants were often unsure whether certain safety functions were available at all, especially for low-salience privacy and audience controls. Plain-language searches in the `Settings' menu commonly failed due to keyword-only matching and narrow indexing, so intent-correct terms (e.g., ``unblock'') did not surface the relevant capability (``block list''). This pattern reflects the \emph{vocabulary problem}—people naturally use diverse, intent-correct words that do not match system labels—forcing incremental probing to learn whether a capability exists \citep{furnas1987vocabulary, teevan2004orienteering}. 

These conditions elevate \emph{learning} costs (time spent hypothesizing and trying variants), spill into \emph{compliance} costs (detours through help centers or off-app searches), and add \emph{psychological} costs (doubt and early abandonment). Accordingly, the remedy is \emph{intent-aware search}: normalize synonyms and near-misses, and surface all semantically relevant controls even when queries do not match system labels~\citep{furnas1987vocabulary, harper2008web}. 

\textbf{(2) Inconsistent labeling.} Identical icons or entry points carried different labels across views, e.g., ``More options,'' ``See more,'' ``Group tools.''  For screen-reader users, these renamings break learned mappings, shift effort from recognition to recall~\citep{Budiu2024RecognitionRecall}, and increase mis-selections and extra gestures. In practice, participants ended up re-learning labels (learning cost), and recovered from avoidable mis-selections—this necessitated extra taps and backtracking that introduced hesitation and slowed the reporting task (compliance cost). 
To address such problems, we recommend that platforms align their web interfaces with the Web Accessibility Initiative's principle of ``Consistent Identification''~\citep{WAI2025_SC324_ConsistentIdentification}.
Using the same label for the same function reduces re-learning overhead and keeps non-visual navigation predictable, echoing broader usability heuristics that emphasize recognition over recall~\citep{Budiu2024RecognitionRecall}.


\textbf{(3) Missing functional cues.} On Facebook, the `Block' control offered no foresight into what activation would entail—and, once invoked, it provided no auditory confirmation that the action was completed; instead, the blocked person's profile was thereafter hidden from the participant's interface, with no indication that this outcome was expected. On X, the pattern was similar: neither `Block' nor `Mute' provided upfront descriptions of their consequences, and explanations appeared only after completing the task, on the confirmation screen. These missing cues widen Norman's \emph{``Gulf of Execution''}—uncertainty about whether the available action matches the user's intent—and \emph{``Gulf of Evaluation''}—insufficient feedback to interpret the new state~\citep{norman2013doet}. A straightforward fix is to provide concise, screen-reader-exposed previews before commitment (e.g., ``Blocking x hides their replies and prevents mentions...'') and immediate, focused auditory confirmations after commitment (e.g., ``You have successfully blocked x...the profile will no linger be visible to you on the platform''), so intent, action, and consequences align in a continuous, understandable flow.

\textbf{(4) Progress ambiguity.}  Participants repeatedly expressed disorientation caused by unclear workflow endpoints and ambiguous mid-task signals. For instance, the lack of a progress indicator in the reporting wizard on Facebook led to premature exits among some participants. According to Tversky and Kahneman's framework of decision-making under uncertainty \citep{tversky1974judgment}, ambiguity amplifies perceived effort, discouraging continued engagement. Thus, ambiguous or incomplete feedback loops in moderation interfaces directly contribute to higher attrition among users with vision loss, intensifying their administrative burden through unnecessary cognitive labor and interaction fatigue. Clear step structures (``Step 2 of 4''), named stages, and immediate confirmations implement the classic visibility-of-system-status heuristic championed by the UX expert Jakob Nielsen~\citep{Nielsen1994, harley2018visibility}.

\textbf{(5) Outcome opacity.} After submitting a report or blocking, users often could not locate results, understand standards applied, or know if/when review would occur. Difficult-to-find status dashboard (called `Support Inbox' on Facebook) and absent outcome trails produce a ``black box'' dynamic—opaque, privately controlled decision systems in which the inputs, operative rules, and rationales are hidden from those affected, blunting accountability and contestation \citep{pasquale2015black}. 
Our investigation found that Facebook's Support Inbox (which can be used for tracking user reports) is hard to locate, often contradicting Help Center guidance and only found via exact keyword search,\footnote{A screen recording of the access path is available at: \url{https://figshare.com/s/b1606fe36094cdb30e3a}.} while X (at the time of study) offered no obvious way to track report outcomes. Such opacity sustains \emph{learning} costs (where is the status kept?), escalates \emph{compliance} costs (repeat submissions), and deepens \emph{psychological} costs (frustration, diminished sense of self-efficacy). A reachable outcomes hub with time-stamped status, rationale, and directions to appeal paths would make user-initiated content moderation more transparent and fair for all users.

In sum, uncertainty around labels, previews, progress, and outcomes did not merely slow down task initiation; it eroded momentum toward completion, making later steps costlier and more fragile. When interfaces did not resolve that uncertainty, participants compensated with \emph{work}—detours, repetitions, and restarts—to keep going. The next subsection examines this labor and its characteristic accumulation patterns.

\subsection{Labor: Inequitable Interaction Work}

Our data show that participants incurred substantial \emph{compliance costs}—the time and effort spent on tedious chores required to execute safety tasks—accumulating through two recurrent mechanisms. First, \textbf{carry-over costs}: when the first step was unclear or hard to reach, extra work \emph{cascaded into later steps}. Second, \textbf{restart costs}: when obvious cues to continue or exit (\textit{Back} or \textit{Cancel} buttons) weren't readily accessible, or when VoiceOver did not focus on any element or focused on the wrong control, people often had to quit their flow and start over. While these mechanisms co-occurred with learning costs (e.g., locating or interpreting controls), our emphasis here is on how interaction work itself accumulated and exacerbated psychological spillovers (hesitation, self-doubt, fatigue).

\textbf{Carry-over costs.}
When the entry point is hard to find or ambiguous, the extra work does not end with discovery; it propagates. Participants who could not reach a reliable entry (e.g., menu items skipped in linear focus order, icons requiring magnification and panning for people with low-vision) accumulated more exploratory gestures upfront and faced longer downstream sequences to reach the same endpoint. The result is a shift from a short, linear path to a branching search-and-verify routine that invites more compliance cost in the form of attention and time.


Even after discovering the correct starting menu icon, safety controls are buried deep within menu trees, and progress requires long chains of micro-actions that redirect attention from judgment to wayfinding. Even on the correct path, reaching a single control can take \emph{dozens} of gestures without search. Unskippable panels and verbose wizards exacerbate the interaction effort required to complete the task. This amount of work turns routine execution into accumulating \emph{compliance costs} while draining users' \emph{psychological reserves}, resulting in fatigue, frustration and attrition.


\textbf{Restart costs} appeared when participants lost their place or the system offered no path backward. Grant selected an incorrect reporting category on X and found no \textit{Back} control on the category selection wizard, forcing him to restart the task. Loss of automatic VoiceOver focus, unintuitive placement of `Back' and `Cancel' buttons in the navigation sequence, and ambiguous progress indicators repeatedly forced participants into abandoning their progress and restarting the task. Restarts are not a single penalty: they erase context, invalidate partial progress, and multiply opportunities for new errors on re-entry. In time-sensitive scenarios (e.g., escalating harassment), these resets translate into missed opportunities for timely intervention, and erode users' willingness to continue reporting, increasing drop-offs. 

 \vspace{8pt}
\noindent
 We read this offloaded coordination work as work extracted from users to keep safety flows moving. Terranova's notion of \emph{free labor}~\citep{terranova2000free} clarifies how platforms benefit from users' enforcement inputs (reports, blocks, classifications) through unpaid work. The burden of this labor is exacerbated for those navigating these controls through inaccessible channels, thus creating an inequitable distribution of the everday burden of staying safe online~\citep{terranova2000free}.
Read together—and situated within the administrative burden of safety work—these frames clarify who pays for the surplus work, why it is inequitable, and how this free labor is unaligned with design-justice commitments to distribute benefits and burdens fairly \citep{CostanzaChock2020}. 


Participants also questioned the fairness~\citep{ma2022m,vaccaro2020end} of shouldering \emph{governance work}: not only executing a flow but translating lived harms into platform taxonomies, interpreting policy language, supplying evidence (e.g., screenshots, timestamps), and then tracking outcomes across scattered surfaces. This aligns with critiques that flagging shifts governance work onto users, who must shoulder the labor of complaint~\citep{Crawford2016Flag}. In practice, this translates into point-of-use ``paperwork'': users must select categories that may not match lived harm, justify choices without previews of consequences, and then monitor opaque channels for outcomes—administrative steps that grow
when platform-side redress infrastructure is weak. While prior work notes that much of the burden of staying safe online still falls on users themselves~\citep{wei2023responsibility,Jhaver2023Personalizing}, our findings show that this burden significantly exacerbates for users with low vision.


To mitigate this inequitable labor, platform designers must shift from a sighted-default paradigm toward proactive and ability-centered design~\citep{Wobbrock2011Ability}. \emph{Proactively surfacing moderation controls with adequate context} directly on posts can substantially reduce navigational complexity. Clear and intuitive navigation structures, explicitly focusing on the first visual element of the page, and readily accessible `Back' and `Cancel' options are essential for streamlining interactions and preventing unnecessary detours. 


Lastly, platforms must fundamentally change their design processes to embrace participatory design methods that meaningfully involve users with vision-related disabilities at every stage of development. This participation would ensure that critical accessibility issues are identified and resolved before becoming entrenched in the infrastructure.  In line with the disability rights principle ``Nothing About Us Without Us''~\citep{CostanzaChock2020}, this means compensating disabled contributors for their expertise and incorporating them as co-designers and decision-makers. Doing so moves responsibility for accessibility upstream, reduces the accumulating administrative burden of safety work, and establishes clearer lines of accountability. These structural commitments foster equity and fairness, advancing disability-justice goals by redistributing design labor more evenly across various stakeholders.

%% file: 07-conclusion.tex
\subsection{Limitations}
While this study provides in-depth qualitative insights into the experiences of blind and low-vision users with content moderation tools on two major social media platforms, we acknowledge certain limitations inherent in our research design and scope. Our investigation, primarily focused on specific user-enacted moderation tasks within Facebook and X, does not represent an exhaustive audit of all accessibility and usability issues across every moderation feature or all social media platforms. The tasks themselves, designed to cover a range of moderation functionalities such as reporting, blocking, and adjusting privacy settings, prompted engagement with tools that participants might not frequently or ever use in their daily routines, potentially not fully capturing organic interaction patterns or motivations. However, we note that the ready usability of such tools becomes critical amid safety concerns, and therefore, tool designs must especially accommodate first-time users. 

The findings reflect a snapshot in time, capturing the state of these platforms, their respective app versions, operating system configurations, and device models encountered during the period of data collection. Given the highly dynamic nature of these digital environments, where interfaces are frequently updated and A/B testing is common, some observed issues may have since been altered, and indeed, we noted instances where specific challenges were not uniformly replicable across researchers' own devices or every participant's encounter. Our primary methodological aim was to highlight the patterns our participants faced and identify systemic barriers during these tasks, recognizing that individual encounters with these issues varied due to differences in personal assistive technology proficiency, adaptive strategies, and the aforementioned platform dynamism. However, our documentation of such systemic barriers, despite the wide popularity and age  of both platforms  suggests a long-standing, industry-wide apathy toward accessibility concerns.

\section{Conclusion}
This study demonstrates how the design of moderation tools on major social media platforms places an unfair burden of safety labor on users with vision impairments. Rectifying this goes beyond technical accessibility --- it requires a genuine commitment to enacting procedural fairness in design. For platforms, this means creating equitable and easily navigable workflows for user-enacted moderation mechanisms, while for researchers, it points toward incorporating accessibility metrics in fairness evaluations of moderation systems and building prototypes that show how online safety is truly achievable for everyone, not just sighted users.